\documentclass[aps,prd,twocolumn,superscriptaddress,floatfix,showpacs,amsmath,amssymb]{revtex4}

\usepackage{afterpage}
\usepackage{subfigure}

\usepackage{graphicx}% Include figure files
\usepackage{dcolumn}% Align table columns on decimal point
\usepackage{bm}% bold math

\usepackage{color}
\usepackage{eso-pic}
\usepackage{epsfig}
\usepackage{rotating}
\usepackage{fancybox}

\newcommand{\ppbar}{$p\bar{p}$}
\newcommand{\Stop}{\tilde{t}}
\newcommand{\antiStop}{\bar{\tilde{t}}}
\newcommand{\chione}[1]{\tilde{\chi}_1^{#1}}
\newcommand{\ttbar}{$t\bar{t}$}
\newcommand{\mufk}{$\mu$+fake}
\newcommand{\pT}{$p_{T}$}
\newcommand{\ET}{$E_\mathrm{T}$}
\newcommand{\MET}{\mbox{$\raisebox{.3ex}{$\not$}$\ET}}

\begin{document}

\title{Search for the supersymmetric partner of the top quark  
in \ppbar~collisions at $\sqrt{s}=$1.96~TeV }

\affiliation{Institute of Physics, Academia Sinica, Taipei, Taiwan 11529, Republic of China}
\affiliation{Argonne National Laboratory, Argonne, Illinois 60439, USA}
\affiliation{University of Athens, 157 71 Athens, Greece}
\affiliation{Institut de Fisica d'Altes Energies, Universitat Autonoma de Barcelona, E-08193, Bellaterra (Barcelona), Spain}
\affiliation{Baylor University, Waco, Texas 76798, USA}
\affiliation{Istituto Nazionale di Fisica Nucleare Bologna, $^{bb}$University of Bologna, I-40127 Bologna, Italy}
\affiliation{Brandeis University, Waltham, Massachusetts 02254, USA}
\affiliation{University of California, Davis, Davis, California 95616, USA}
\affiliation{University of California, Los Angeles, Los Angeles, California 90024, USA}
\affiliation{Instituto de Fisica de Cantabria, CSIC-University of Cantabria, 39005 Santander, Spain}
\affiliation{Carnegie Mellon University, Pittsburgh, Pennsylvania 15213, USA}
\affiliation{Enrico Fermi Institute, University of Chicago, Chicago, Illinois 60637, USA}
\affiliation{Comenius University, 842 48 Bratislava, Slovakia; Institute of Experimental Physics, 040 01 Kosice, Slovakia}
\affiliation{Joint Institute for Nuclear Research, RU-141980 Dubna, Russia}
\affiliation{Duke University, Durham, North Carolina 27708, USA}
\affiliation{Fermi National Accelerator Laboratory, Batavia, Illinois 60510, USA}
\affiliation{University of Florida, Gainesville, Florida 32611, USA}
\affiliation{Laboratori Nazionali di Frascati, Istituto Nazionale di Fisica Nucleare, I-00044 Frascati, Italy}
\affiliation{University of Geneva, CH-1211 Geneva 4, Switzerland}
\affiliation{Glasgow University, Glasgow G12 8QQ, United Kingdom}
\affiliation{Harvard University, Cambridge, Massachusetts 02138, USA}
\affiliation{Division of High Energy Physics, Department of Physics, University of Helsinki and Helsinki Institute of Physics, FIN-00014, Helsinki, Finland}
\affiliation{University of Illinois, Urbana, Illinois 61801, USA}
\affiliation{The Johns Hopkins University, Baltimore, Maryland 21218, USA}
\affiliation{Institut f\"{u}r Experimentelle Kernphysik, Karlsruhe Institute of Technology, D-76131 Karlsruhe, Germany}
\affiliation{Center for High Energy Physics: Kyungpook National University, Daegu 702-701, Korea; Seoul National University, Seoul 151-742, Korea; Sungkyunkwan University, Suwon 440-746, Korea; Korea Institute of Science and Technology Information, Daejeon 305-806, Korea; Chonnam National University, Gwangju 500-757, Korea; Chonbuk National University, Jeonju 561-756, Korea}
\affiliation{Ernest Orlando Lawrence Berkeley National Laboratory, Berkeley, California 94720, USA}
\affiliation{University of Liverpool, Liverpool L69 7ZE, United Kingdom}
\affiliation{University College London, London WC1E 6BT, United Kingdom}
\affiliation{Centro de Investigaciones Energeticas Medioambientales y Tecnologicas, E-28040 Madrid, Spain}
\affiliation{Massachusetts Institute of Technology, Cambridge, Massachusetts 02139, USA}
\affiliation{Institute of Particle Physics: McGill University, Montr\'{e}al, Qu\'{e}bec, Canada H3A~2T8; Simon Fraser University, Burnaby, British Columbia, Canada V5A~1S6; University of Toronto, Toronto, Ontario, Canada M5S~1A7; and TRIUMF, Vancouver, British Columbia, Canada V6T~2A3}
\affiliation{University of Michigan, Ann Arbor, Michigan 48109, USA}
\affiliation{Michigan State University, East Lansing, Michigan 48824, USA}
\affiliation{Institution for Theoretical and Experimental Physics, ITEP, Moscow 117259, Russia}
\affiliation{University of New Mexico, Albuquerque, New Mexico 87131, USA}
\affiliation{Northwestern University, Evanston, Illinois 60208, USA}
\affiliation{The Ohio State University, Columbus, Ohio 43210, USA}
\affiliation{Okayama University, Okayama 700-8530, Japan}
\affiliation{Osaka City University, Osaka 588, Japan}
\affiliation{University of Oxford, Oxford OX1 3RH, United Kingdom}
\affiliation{Istituto Nazionale di Fisica Nucleare, Sezione di Padova-Trento, $^{cc}$University of Padova, I-35131 Padova, Italy}
\affiliation{LPNHE, Universite Pierre et Marie Curie/IN2P3-CNRS, UMR7585, Paris, F-75252 France}
\affiliation{University of Pennsylvania, Philadelphia, Pennsylvania 19104, USA}
\affiliation{Istituto Nazionale di Fisica Nucleare Pisa, $^{dd}$University of Pisa, $^{ee}$University of Siena and $^{ff}$Scuola Normale Superiore, I-56127 Pisa, Italy}
\affiliation{University of Pittsburgh, Pittsburgh, Pennsylvania 15260, USA}
\affiliation{Purdue University, West Lafayette, Indiana 47907, USA}
\affiliation{University of Rochester, Rochester, New York 14627, USA}
\affiliation{The Rockefeller University, New York, New York 10065, USA}
\affiliation{Istituto Nazionale di Fisica Nucleare, Sezione di Roma 1, $^{gg}$Sapienza Universit\`{a} di Roma, I-00185 Roma, Italy}

\affiliation{Rutgers University, Piscataway, New Jersey 08855, USA}
\affiliation{Texas A\&M University, College Station, Texas 77843, USA}
\affiliation{Istituto Nazionale di Fisica Nucleare Trieste/Udine, I-34100 Trieste, $^{hh}$University of Trieste/Udine, I-33100 Udine, Italy}
\affiliation{University of Tsukuba, Tsukuba, Ibaraki 305, Japan}
\affiliation{Tufts University, Medford, Massachusetts 02155, USA}
\affiliation{Waseda University, Tokyo 169, Japan}
\affiliation{Wayne State University, Detroit, Michigan 48201, USA}
\affiliation{University of Wisconsin, Madison, Wisconsin 53706, USA}
\affiliation{Yale University, New Haven, Connecticut 06520, USA}
\author{T.~Aaltonen}
\affiliation{Division of High Energy Physics, Department of Physics, University of Helsinki and Helsinki Institute of Physics, FIN-00014, Helsinki, Finland}
\author{B.~\'{A}lvarez~Gonz\'{a}lez$^v$}
\affiliation{Instituto de Fisica de Cantabria, CSIC-University of Cantabria, 39005 Santander, Spain}
\author{S.~Amerio}
\affiliation{Istituto Nazionale di Fisica Nucleare, Sezione di Padova-Trento, $^{cc}$University of Padova, I-35131 Padova, Italy}

\author{D.~Amidei}
\affiliation{University of Michigan, Ann Arbor, Michigan 48109, USA}
\author{A.~Anastassov}
\affiliation{Northwestern University, Evanston, Illinois 60208, USA}
\author{A.~Annovi}
\affiliation{Laboratori Nazionali di Frascati, Istituto Nazionale di Fisica Nucleare, I-00044 Frascati, Italy}
\author{J.~Antos}
\affiliation{Comenius University, 842 48 Bratislava, Slovakia; Institute of Experimental Physics, 040 01 Kosice, Slovakia}
\author{G.~Apollinari}
\affiliation{Fermi National Accelerator Laboratory, Batavia, Illinois 60510, USA}
\author{J.A.~Appel}
\affiliation{Fermi National Accelerator Laboratory, Batavia, Illinois 60510, USA}
\author{A.~Apresyan}
\affiliation{Purdue University, West Lafayette, Indiana 47907, USA}
% ...
\collaboration{CDF Collaboration\footnote{With visitors from $^a$University of Massachusetts Amherst, Amherst, Massachusetts 01003,
$^b$Istituto Nazionale di Fisica Nucleare, Sezione di Cagliari, 09042 Monserrato (Cagliari), Italy,
$^c$University of California Irvine, Irvine, CA  92697,
$^d$University of California Santa Barbara, Santa Barbara, CA 93106
$^e$University of California Santa Cruz, Santa Cruz, CA  95064,
$^f$CERN,CH-1211 Geneva, Switzerland,
$^g$Cornell University, Ithaca, NY  14853,
$^h$University of Cyprus, Nicosia CY-1678, Cyprus,
$^i$University College Dublin, Dublin 4, Ireland,
$^j$University of Fukui, Fukui City, Fukui Prefecture, Japan 910-0017,
$^k$Universidad Iberoamericana, Mexico D.F., Mexico,
$^l$Iowa State University, Ames, IA  50011,
$^m$University of Iowa, Iowa City, IA  52242,
$^n$Kinki University, Higashi-Osaka City, Japan 577-8502,
$^o$Kansas State University, Manhattan, KS 66506,
$^p$University of Manchester, Manchester M13 9PL, England,
$^q$Queen Mary, University of London, London, E1 4NS, England,
$^r$Muons, Inc., Batavia, IL 60510,
$^s$Nagasaki Institute of Applied Science, Nagasaki, Japan,
$^t$National Research Nuclear University, Moscow, Russia,
$^u$University of Notre Dame, Notre Dame, IN 46556,
$^v$Universidad de Oviedo, E-33007 Oviedo, Spain,
$^w$Texas Tech University, Lubbock, TX  79609,
$^x$IFIC(CSIC-Universitat de Valencia), 56071 Valencia, Spain,
$^y$Universidad Tecnica Federico Santa Maria, 110v Valparaiso, Chile,
$^z$University of Virginia, Charlottesville, VA  22906,
$^{aa}$Yarmouk University, Irbid 211-63, Jordan,
$^{ii}$On leave from J.~Stefan Institute, Ljubljana, Slovenia,
}}
\noaffiliation

\date{\today}

\begin{abstract}

We present a search for the lightest supersymmetric partner of the top quark
in proton-antiproton collisions
at a center-of-mass energy $\sqrt{s}=1.96$~TeV. This search was conducted
within the framework of the $R$-parity conserving minimal supersymmetric extension
of the standard model, assuming the stop decays dominantly to a lepton, a sneutrino,
and a bottom quark.
We searched for events with two oppositely-charged  
leptons, at least one jet, and missing transverse energy in a data sample corresponding 
to an integrated luminosity of 1 fb$^{-1}$ collected by the CDF experiment.  
No significant evidence of a stop quark signal
was found. Exclusion limits at 95\% confidence level in the stop quark versus 
sneutrino mass plane are set. Stop quark masses up to 180~GeV/$c^2$ are excluded  
for sneutrino masses around 45~GeV/$c^2$, and 
sneutrino masses up to 116~GeV/$c^2$ are excluded for 
stop quark masses around 150~GeV/$c^2$.

\end{abstract}

% activate the following line for publication

\maketitle

\newpage

\section{INTRODUCTION}

The minimal supersymmetric standard model \cite{MSSM} (MSSM) was introduced
to solve several problems that arise in the
standard model (SM). These include: the hierarchy problem that requires the
fine-tuning of theoretical parameters in order to obtain cancellation of large
quantum corrections to the Higgs mass; the lack of convergence of the strong, weak, and
electromagnetic gauge couplings at the grand-unification energy scale; and the lack of
a dark matter candidate.

The MSSM assigns a new bosonic counterpart to each SM fermion and likewise
a fermionic superpartner to each SM boson. This results in scalar partners $\tilde{q}_{L}$
and $\tilde{q}_{R}$ to the SM quark helicity states $q_L$ and $q_R$.
There can be two supersymmetric mass eigenstates for each supersymmetric quark (squark)
corresponding to the two fermionic degrees-of-freedom of the SM quark.
The supersymmetric scalar top quark (stop) mass eigenstates
$\tilde{t}_{1}$ and $\tilde{t}_{2}$ are rotated relative
to $\tilde{t}_{L}$ and $\tilde{t}_{R}$ by a mixing angle $\theta_{\tilde{t}}$.
In some models \cite{split}, $m_{\tilde{t}_{1,2}} \approx m_t$.
The large mass of the top quark and the corresponding large value of the top-to-Higgs coupling constant
may lead to a large splitting between $m_{\tilde{t}_{1}}$ and $m_{\tilde{t}_{2}}$.
Consequently the lower mass stop quark eigenstate is expected
to be the lightest of all the squarks, with a mass even below the
top quark, making its detection at the Tevatron a realistic
possibility.

The MSSM possesses a new conserved quantity
called $R$-parity ($R_\mathrm{p}$), defined as $R_\mathrm{p}~=~(-1)^\mathrm{3(B-L)+2S}$, where
$B$ is the baryon number, $L$ is the lepton number, and $S$ is the spin.
As a consequence, the lightest supersymmetric particle (LSP) must
be stable, and is a dark matter candidate.  Because the initial state of
$p\bar{p}$ collisions has $R_\mathrm{p} = +1$ and supersymmetric particles have $R_\mathrm{p} = -1$,
supersymmetric particles must be pair-produced. At the Tevatron
stop quarks
are expected to be produced primarily through gluon-gluon ($gg$) fusion and
quark-antiquark ($q\bar{q}$) annihilation,
with $gg$ fusion dominant at low stop masses ($<100$~GeV/$c^2$) and $q\bar{q}$ annihilation
dominant at higher stop masses \cite{beenakker}.

The produced stop ($\Stop$) quarks can decay via several possible channels, depending on
the masses of the particles involved. Two-body decays include
$\Stop \rightarrow t\chione{0}$,
$\Stop \rightarrow b\chione{+}$, and
$\Stop \rightarrow c\tilde{\chi}_{1}^0$
where $\chione{0}$ is the lightest neutralino and $\chione{+}$ the lightest chargino.
These decays
may not be  kinematically possible for a light stop or they may be suppressed by higher
order diagrams.
They are also constrained by existing limits \cite{LEP,CDFD0}.
Possible three-body decays include
$\Stop \rightarrow W^+  b\chione{0}$,
$\Stop \rightarrow b \tilde{l}^+ \nu_l$, and
$\Stop \rightarrow b l^+ \tilde{\nu}_{l}$
where $\tilde{l}$ is the supersymmetric lepton
and $\tilde{\nu}_{l}$ the supersymmetric neutrino.
Limits on the supersymmetric lepton and neutralino
masses from experiments at the large electron-positron collider (LEP) \cite{LEP} restrict
the range of stop masses available to the first two decay modes.

The decay $\Stop \rightarrow b l^+ \tilde{\nu}_{l}$, which proceeds via a
virtual chargino, is the subject of
this analysis. We assume the branching ratio for this decay mode
is 100\% and that electrons, muons,
and taus are equally likely decay products. While electrons and muons
are detected directly, taus are only included opportunistically in this analysis
through their decays into electrons and muons.
We also assume that the supersymmetric neutrino decays neutrally
into the LSP (or is the LSP), thus escaping
undetected and leading to missing transverse energy ($\not$$E_\mathrm{T}$) \cite{met} in the detector.
Since stop quarks are produced in pairs, we search for events with two
opposite-charge leptons ($ee$, $e\mu$, $\mu\mu$), $\not$$E_\mathrm{T}$,
and at least one hadronic jet.

Previous searches  at LEP and at the Tevatron \cite{LEP, CDFD0}
for the stop quark  using the same topology as this analysis have produced negative results.
These analyses have set $95\%$ confidence level exclusion limits in the stop-sneutrino mass plane.
This paper extends the earlier CDF results based on 107 pb$^{-1}$ of data at $\sqrt{s}=1.8$~TeV
to 1 fb$^{-1}$ at $\sqrt{s}=1.96$~TeV.

The structure of this
paper is as follows: Section II details the detector and the data set; Sections III
describes the background estimation; Section IV the
signal predictions and systematic uncertainties; Section V
event pre-selection and control samples;
Section VI explains the optimization of event selection cuts;
and Section VII presents the results and conclusions.

\begin{figure*}[htpb]
\centerline{\includegraphics[width=0.50\textwidth]{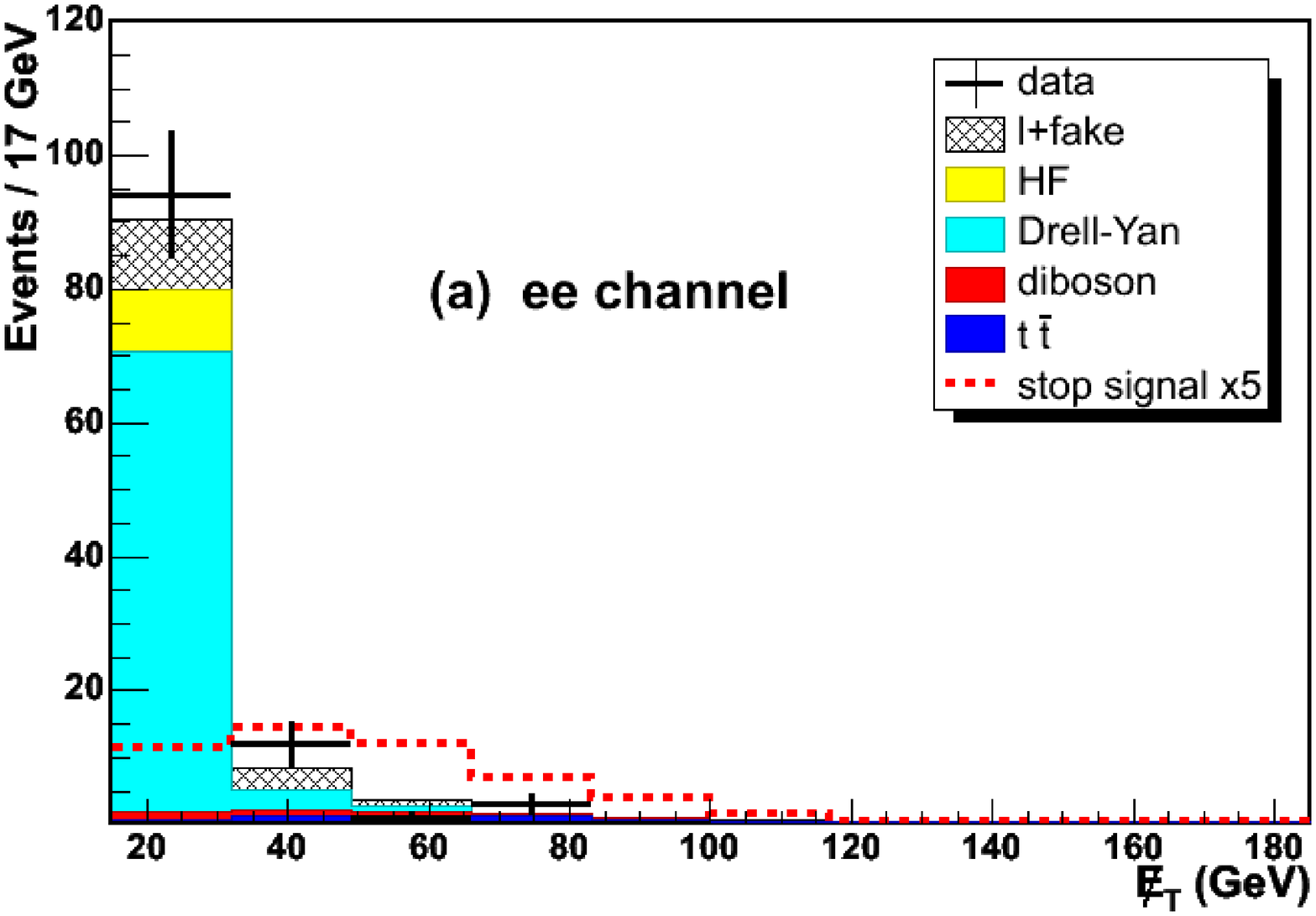}
   \hspace*{-.3cm}
   \includegraphics[width=0.50\textwidth]{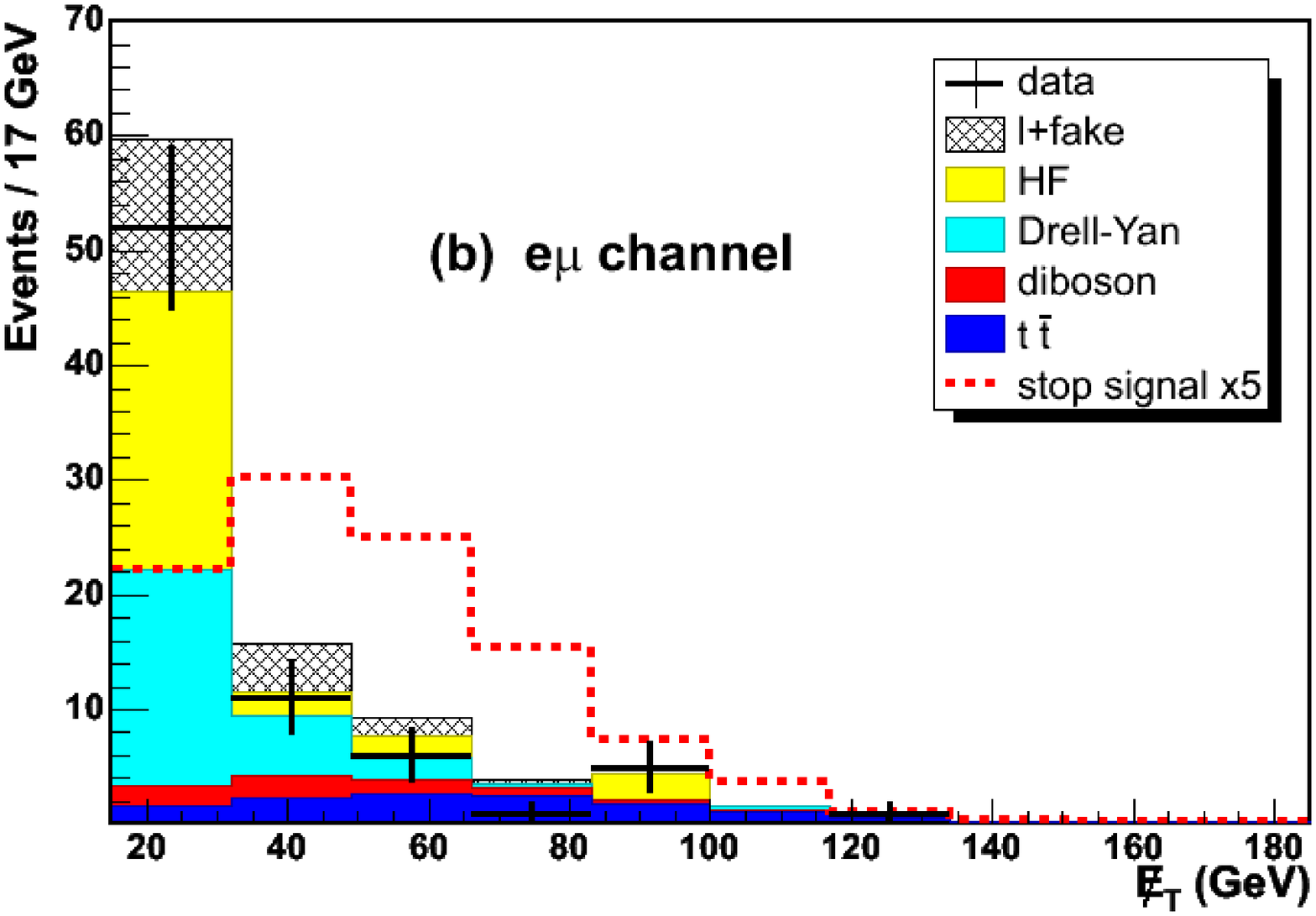}
 }
\centerline{\includegraphics[width=0.50\textwidth]{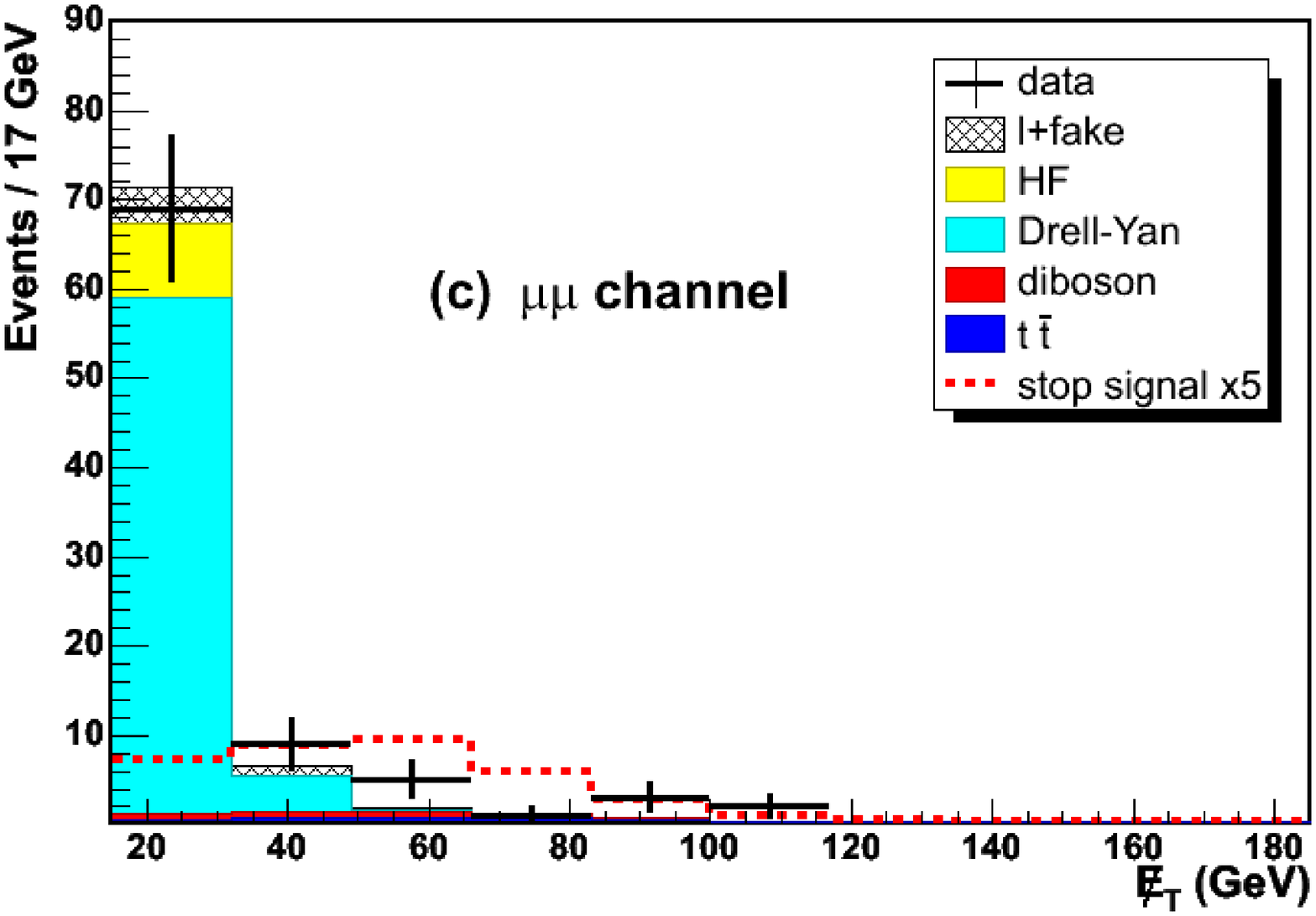}
   \hspace*{-.3cm}
   \includegraphics[width=0.50\textwidth]{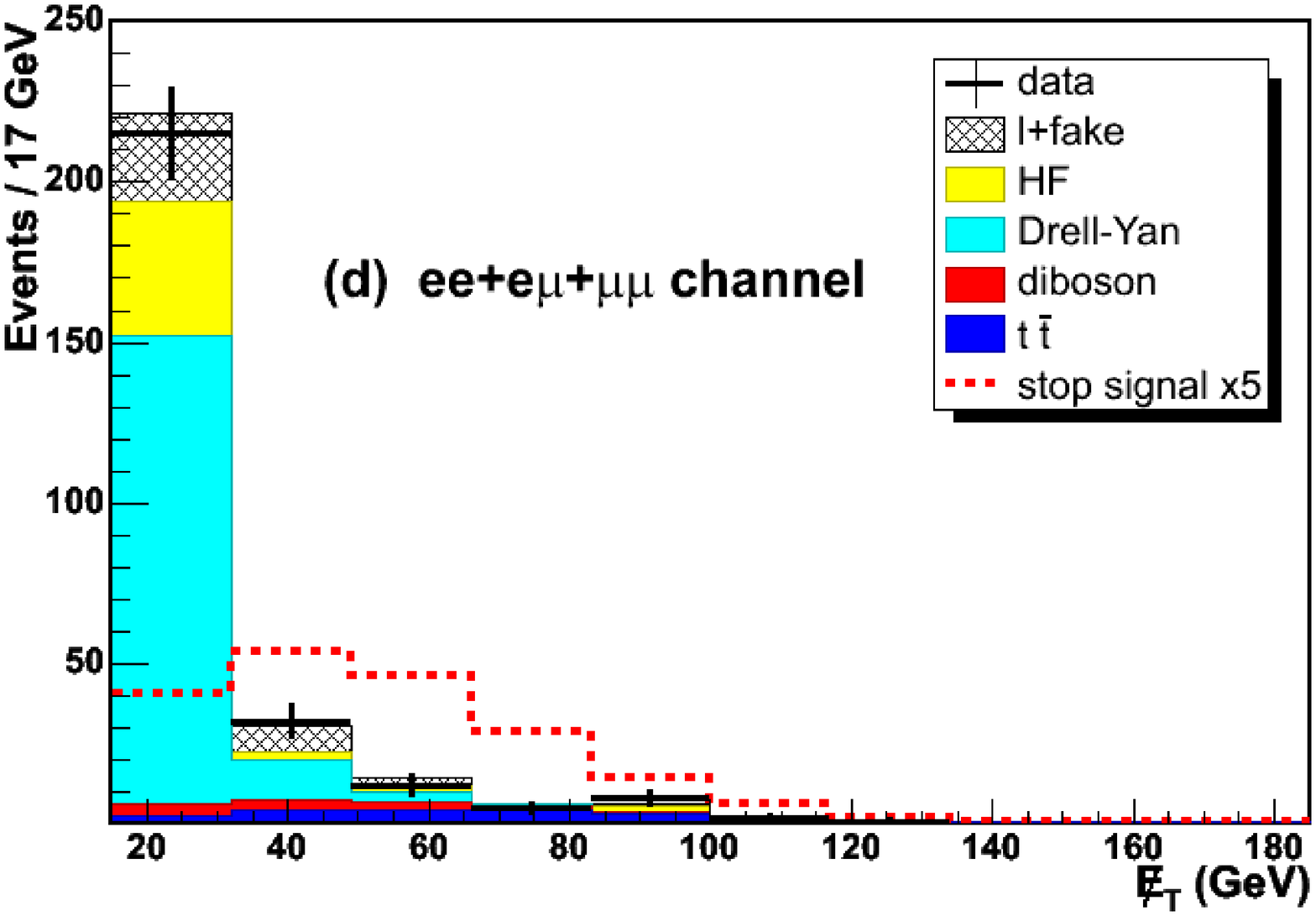}
 }
   \caption{\label{fig:MET_presig}
$\not$$E_\mathrm{T}$ distributions in the pre-signal region for (a) the $ee$ channel,
(b) the $e\mu$ channel,
(c) the $\mu\mu$ channel, and (d) the three channels combined. Data are shown as the
points with  error bars (statistical only). Shown as stacked histograms are the
backgrounds arising from misidentified hadrons and decays-in-flight ($l$+fake),
$b \bar b$ and $c \bar c$ (HF), DY, dibosons, and $t \bar t$.
For reference, the expected signal for $(m_{\tilde{t}},m_{\tilde{\nu}})$ = (150,75)~GeV/$c^2$,
multiplied by five, is shown as the dashed line. }
\end{figure*}

\begin{figure*}[htpb]
\centerline{\includegraphics[width=0.50\textwidth]{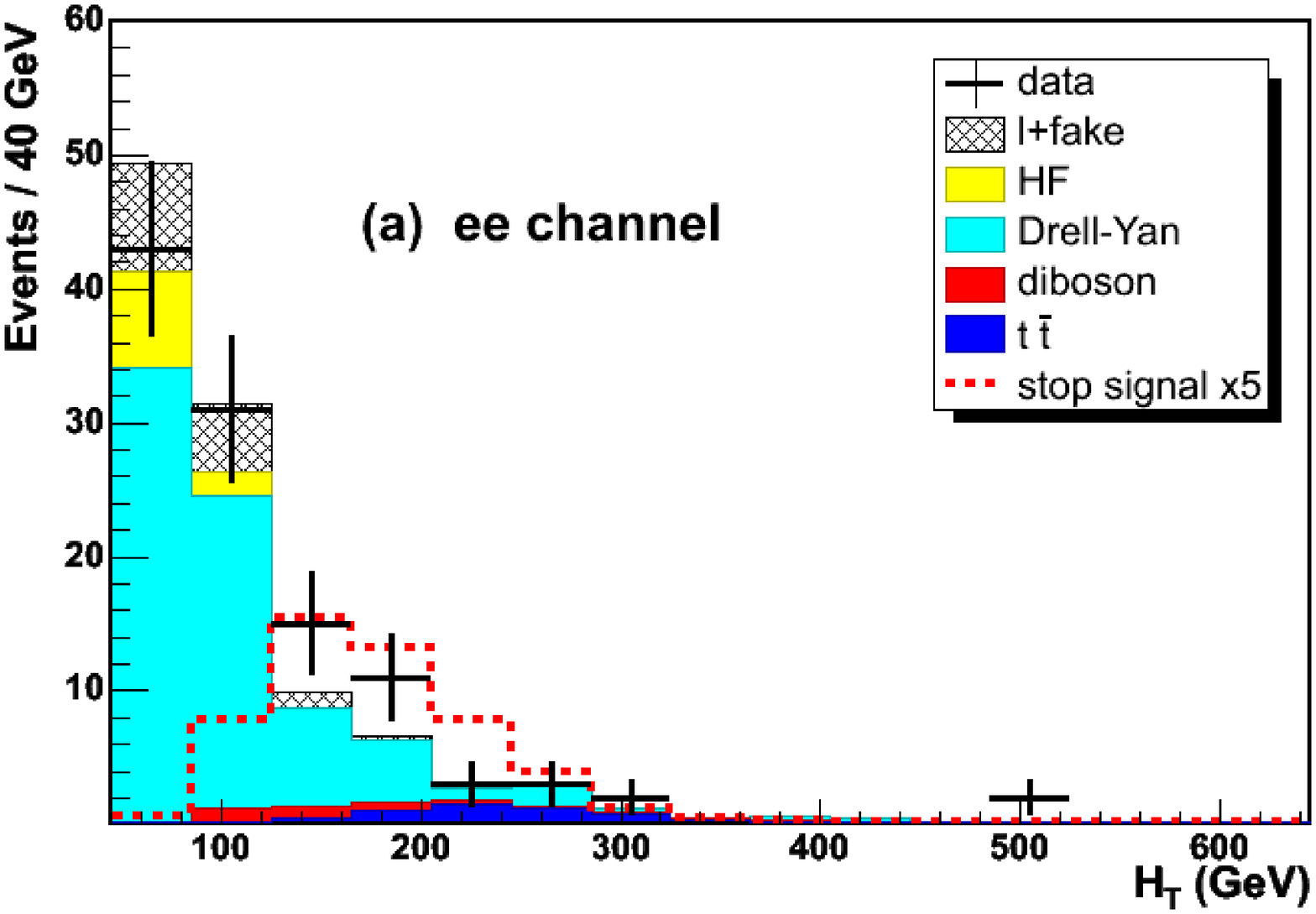}
   \hspace*{-.3cm}
   \includegraphics[width=0.50\textwidth]{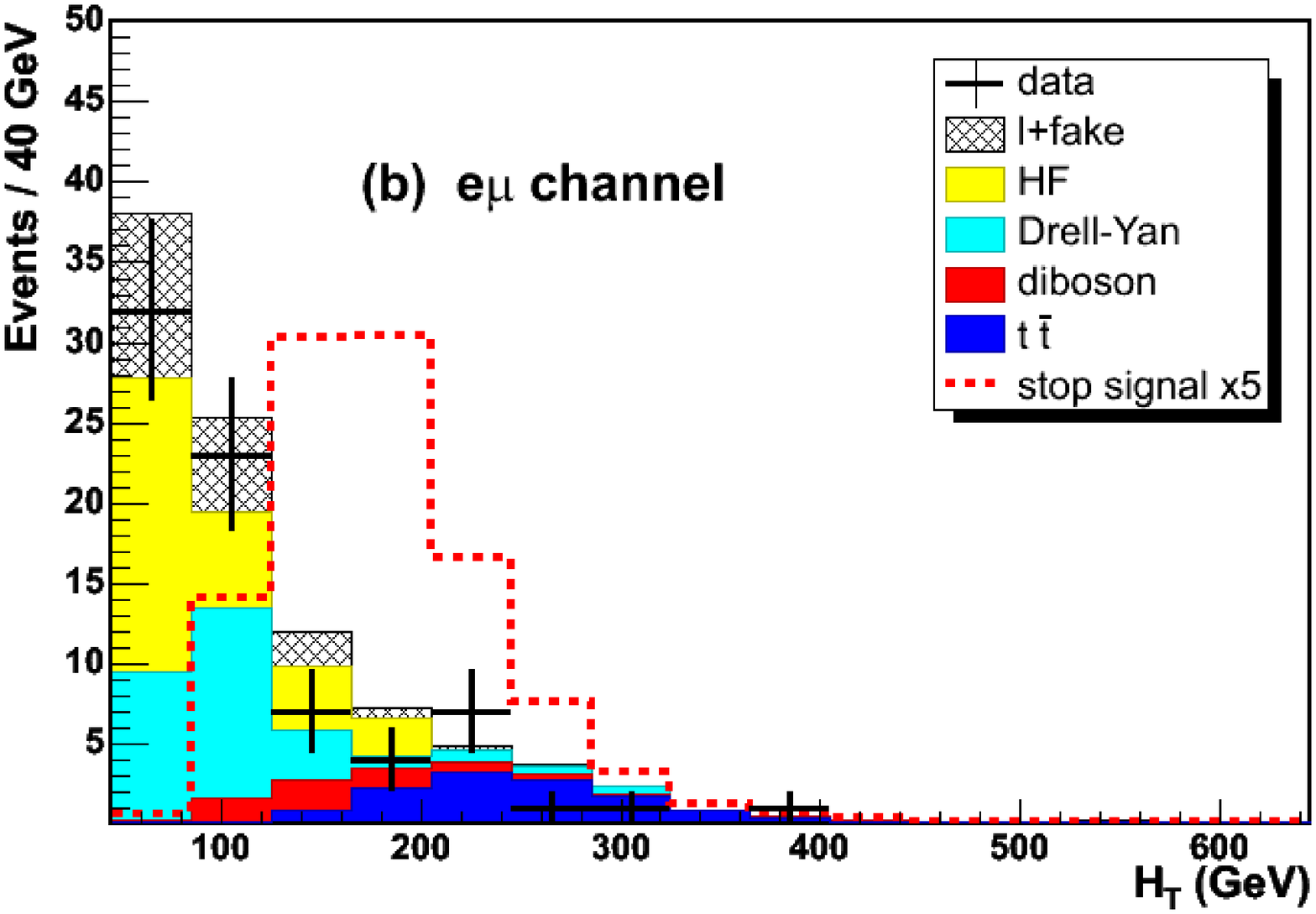}
 }
\centerline{\includegraphics[width=0.50\textwidth]{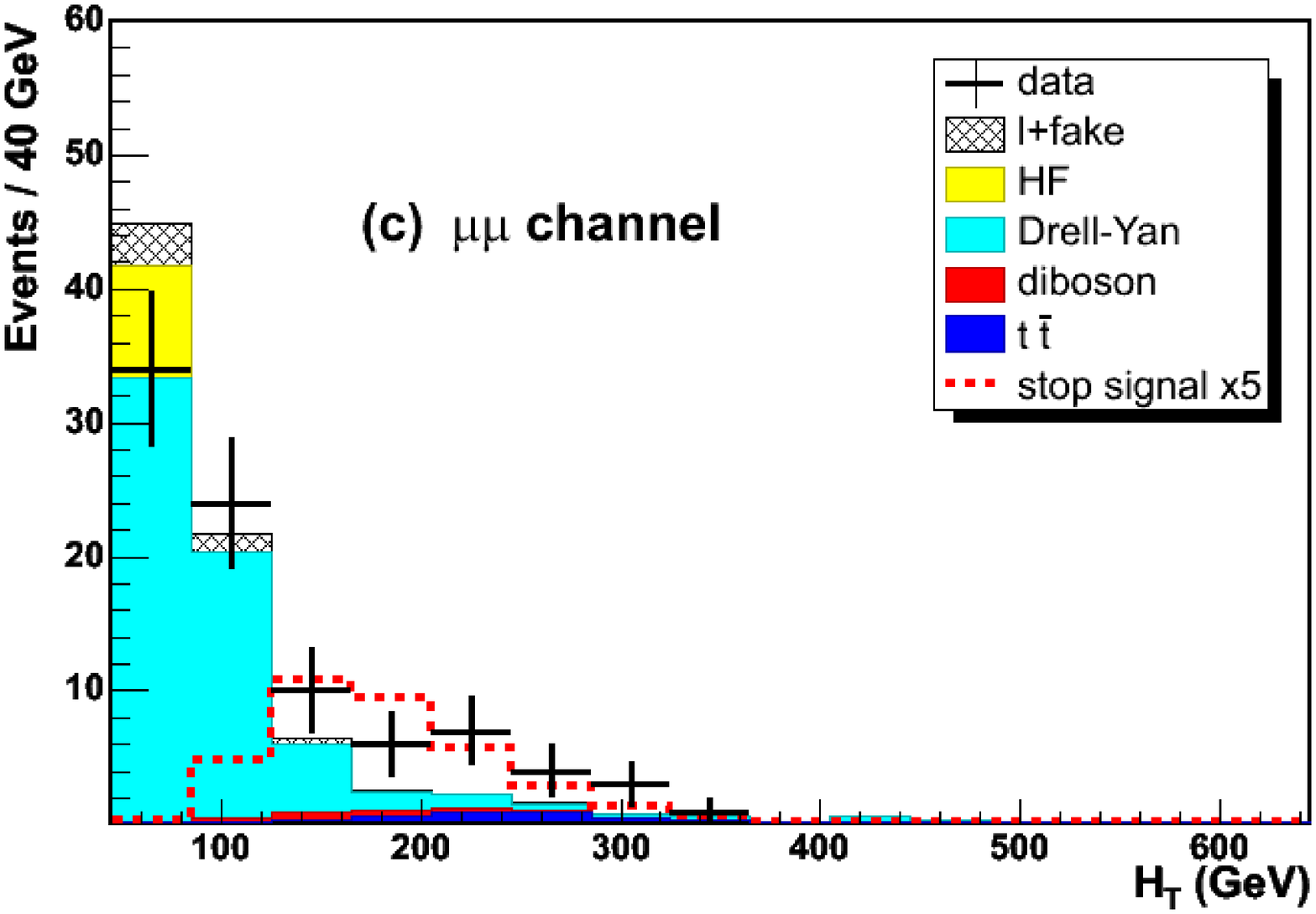}
   \hspace*{-.3cm}
   \includegraphics[width=0.50\textwidth]{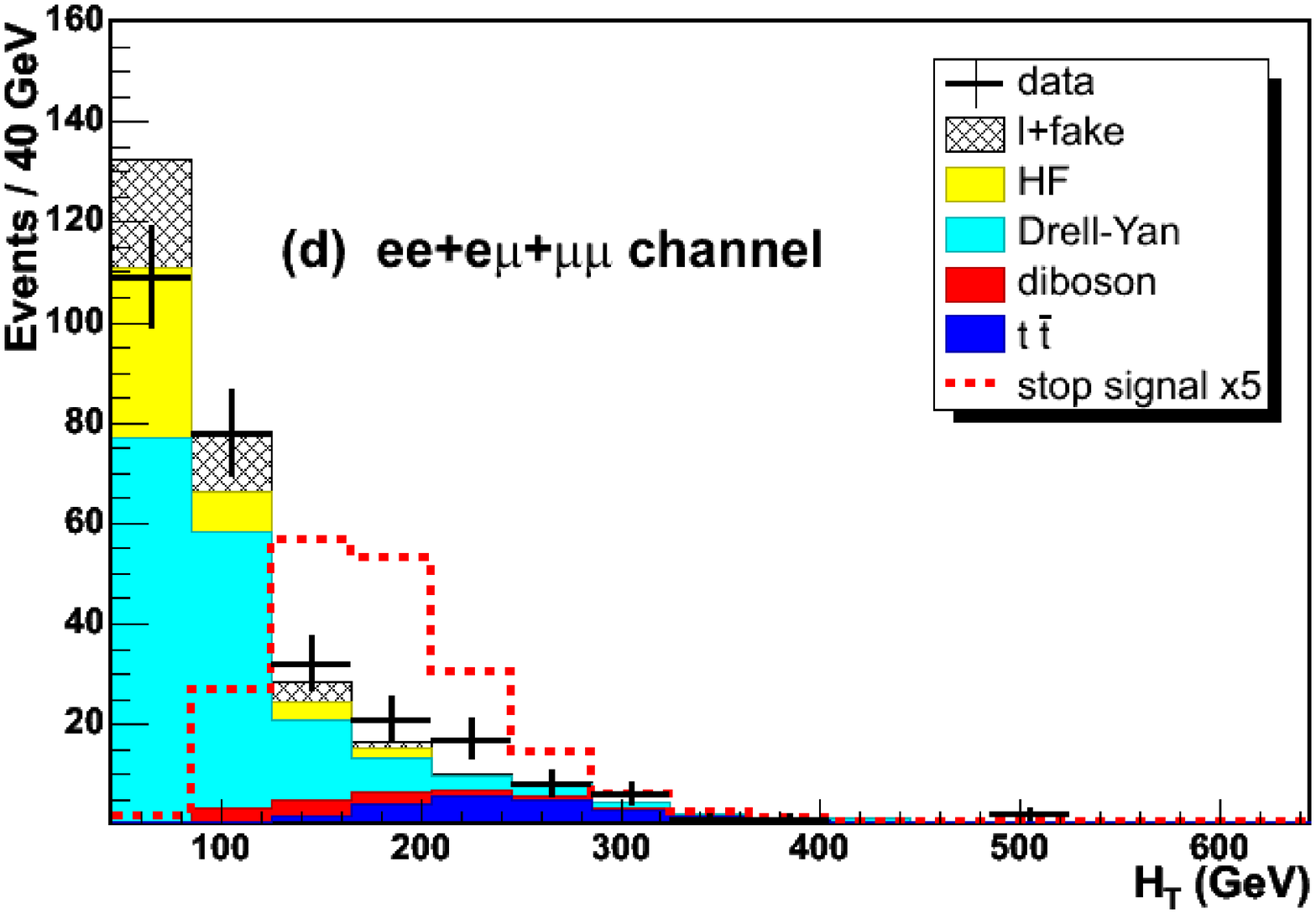}
 }
   \caption{\label{fig:HT_presig}
$H_\mathrm{T}$ distributions in the pre-signal region, shown as in Fig.~\ref{fig:MET_presig}. }
\end{figure*}

\begin{figure*}[htpb]
\centerline{\includegraphics[width=0.50\textwidth]{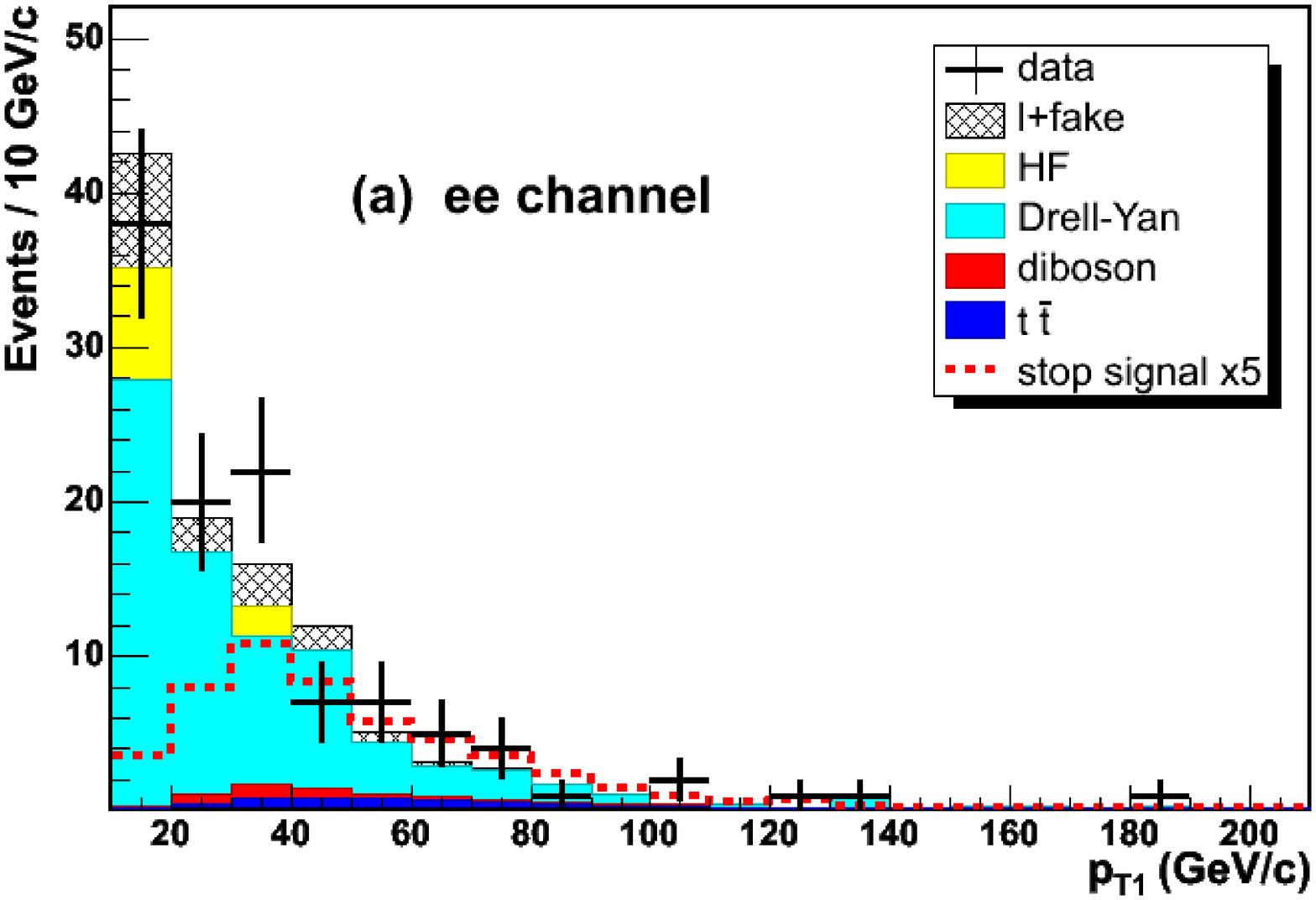}
   \hspace*{-.3cm}
   \includegraphics[width=0.50\textwidth]{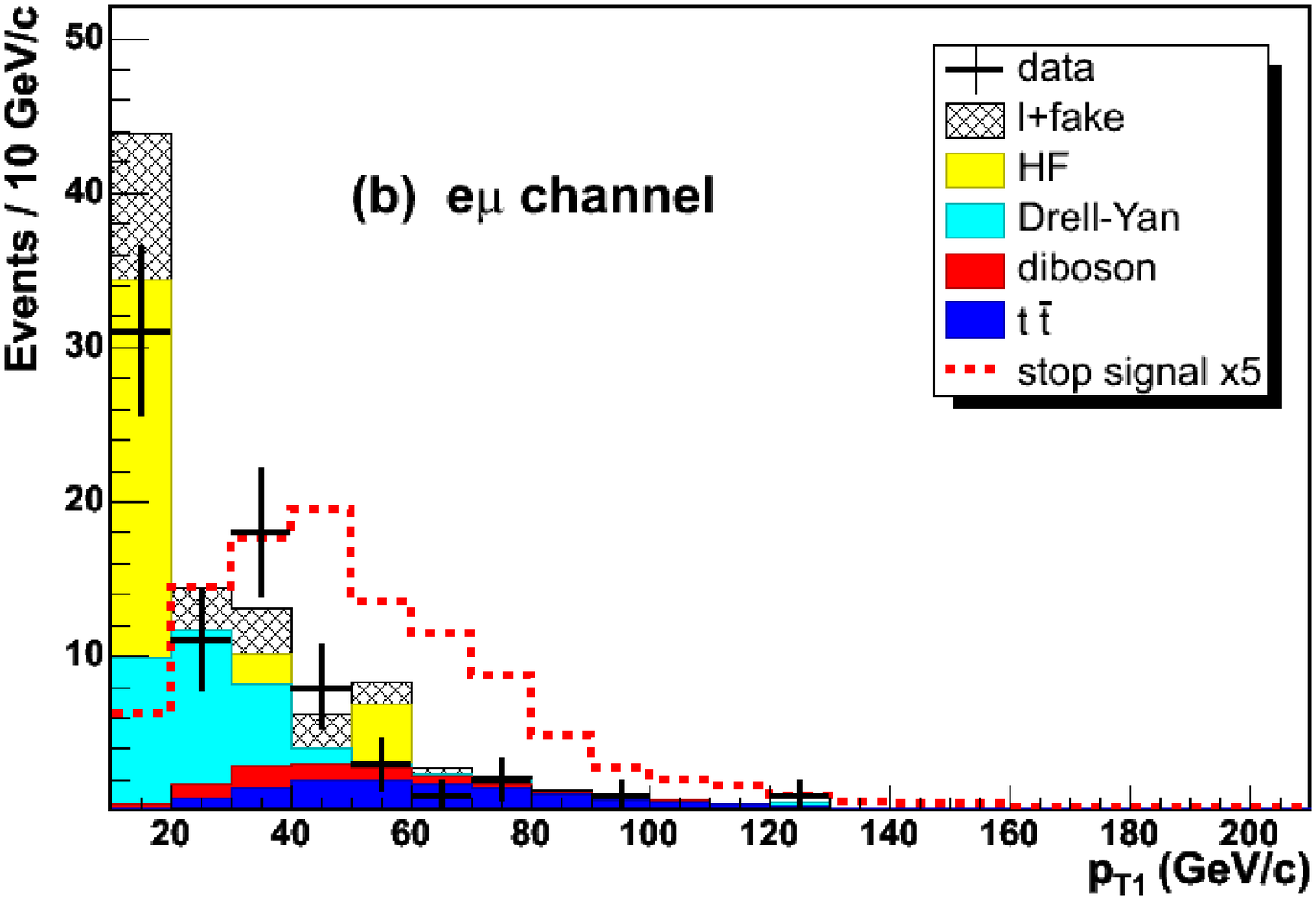}
 }
\centerline{\includegraphics[width=0.50\textwidth]{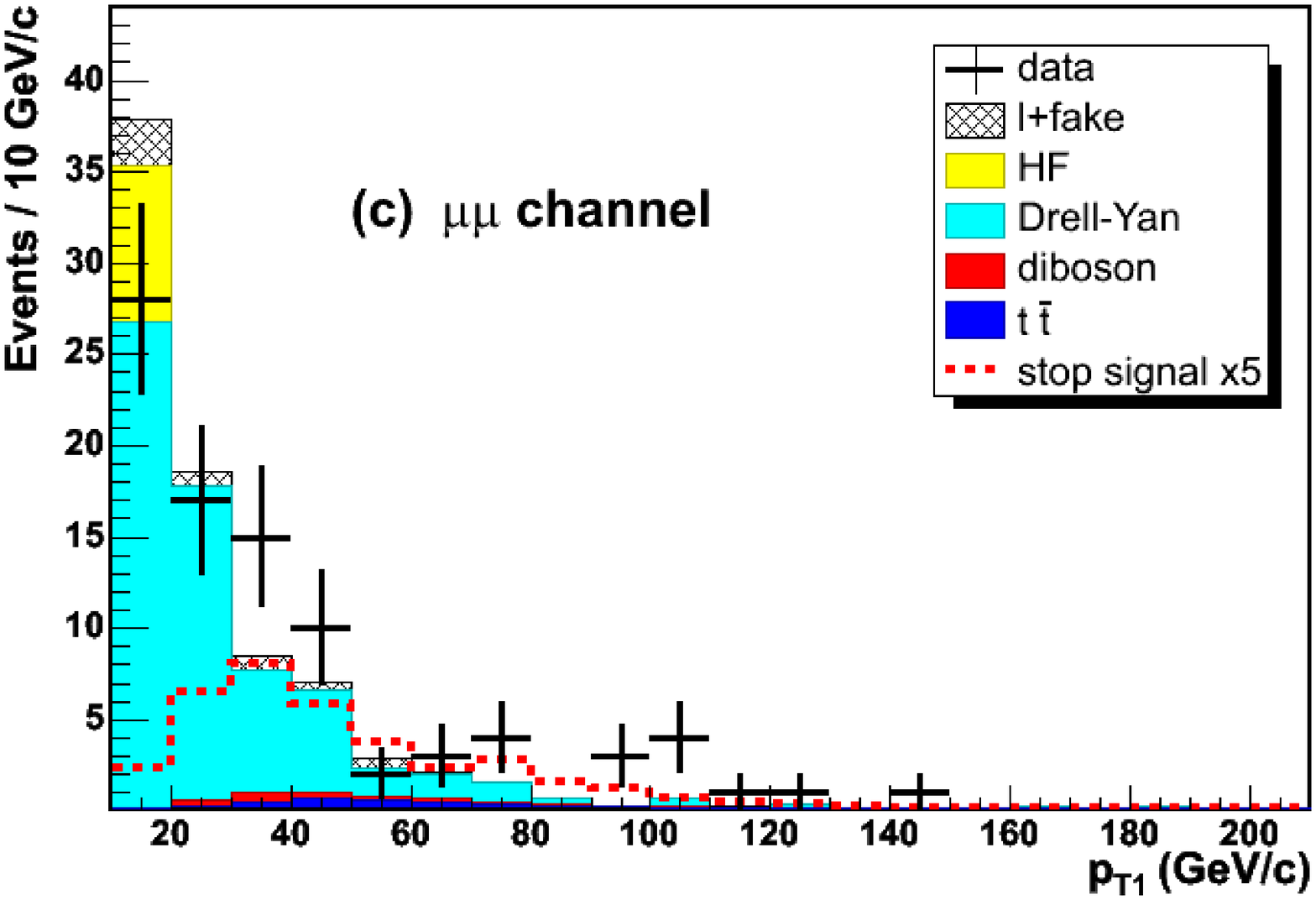}
   \hspace*{-.3cm}
   \includegraphics[width=0.50\textwidth]{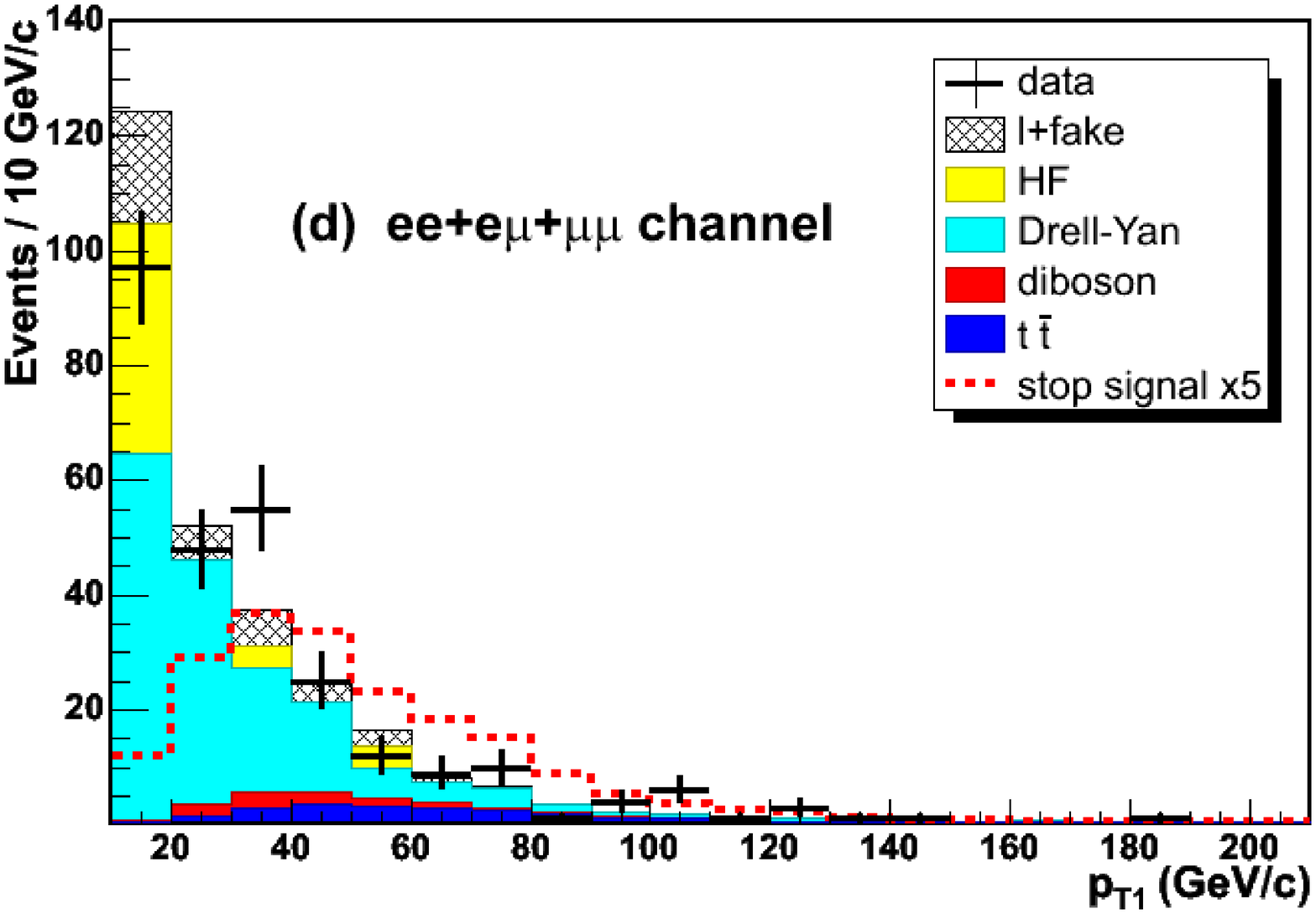}
 }
   \caption{\label{fig:PT1_presig}
$p_\mathrm{T}$ distributions of the highest $p_\mathrm{T}$ lepton in the pre-signal region,
shown as in Fig.~\ref{fig:MET_presig}. }
\end{figure*}

\begin{figure*}[htpb]
\centerline{\includegraphics[width=0.50\textwidth]{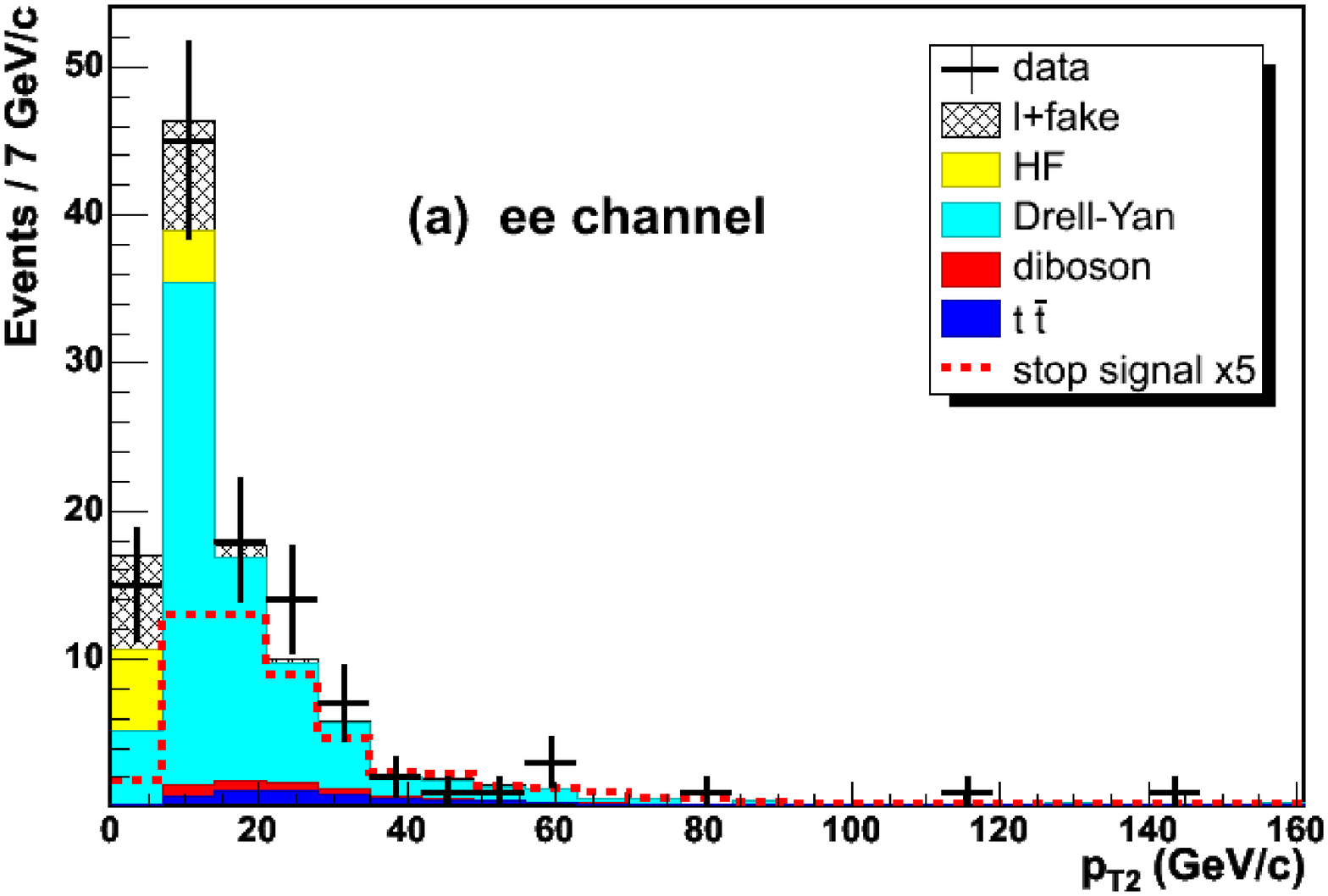}
   \hspace*{-.3cm}
   \includegraphics[width=0.50\textwidth]{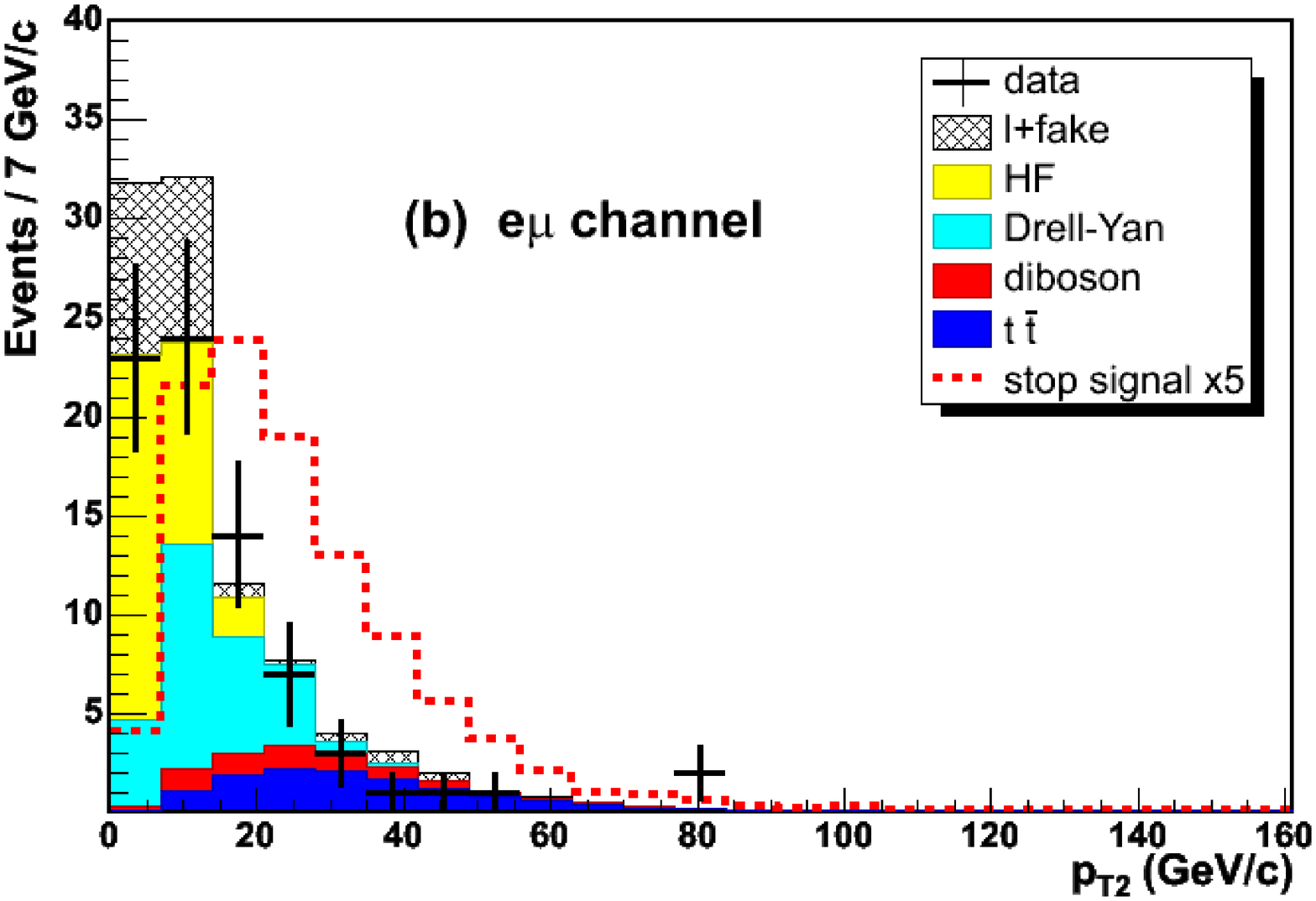}
 }
\centerline{\includegraphics[width=0.50\textwidth]{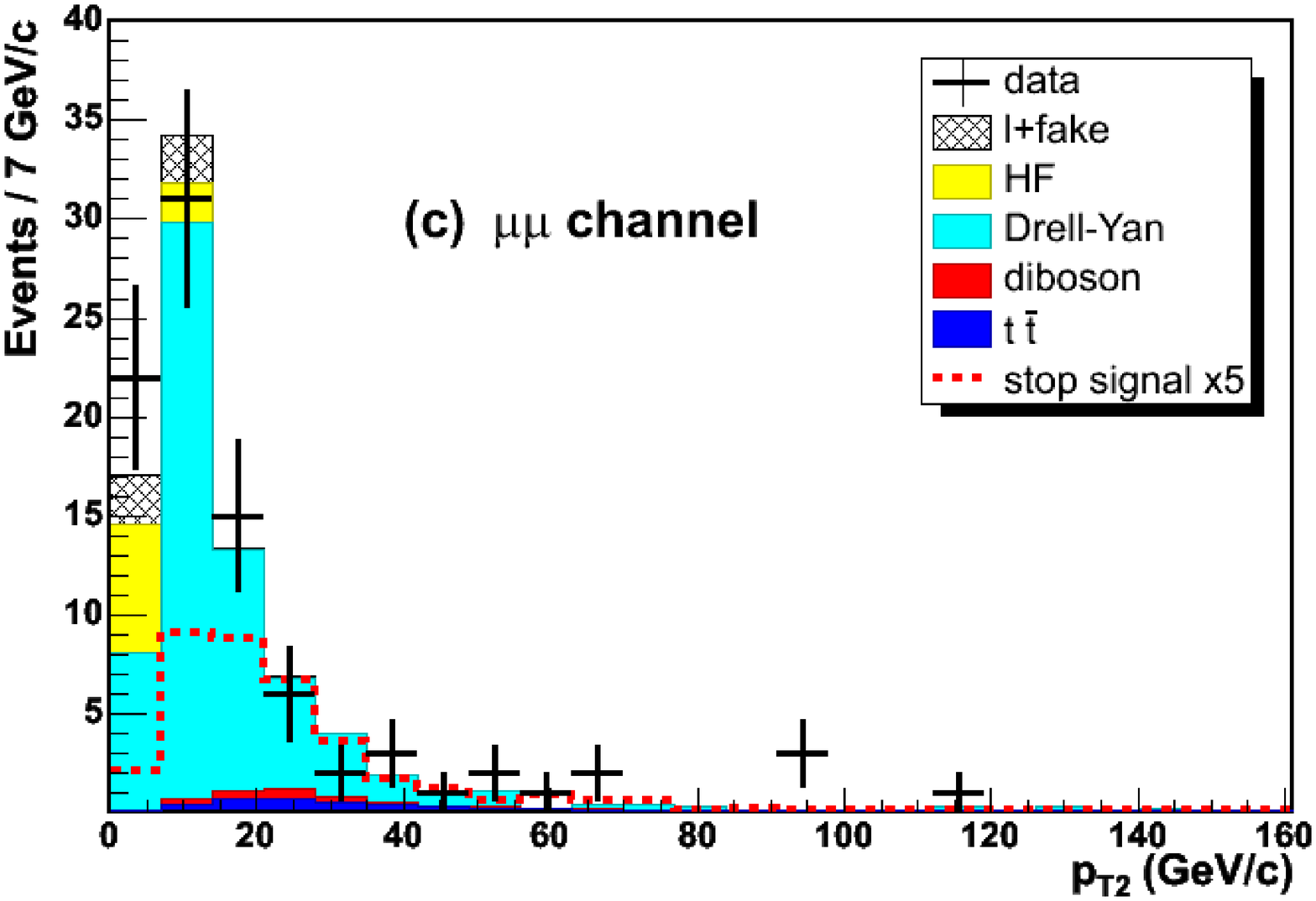}
   \hspace*{-.3cm}
   \includegraphics[width=0.50\textwidth]{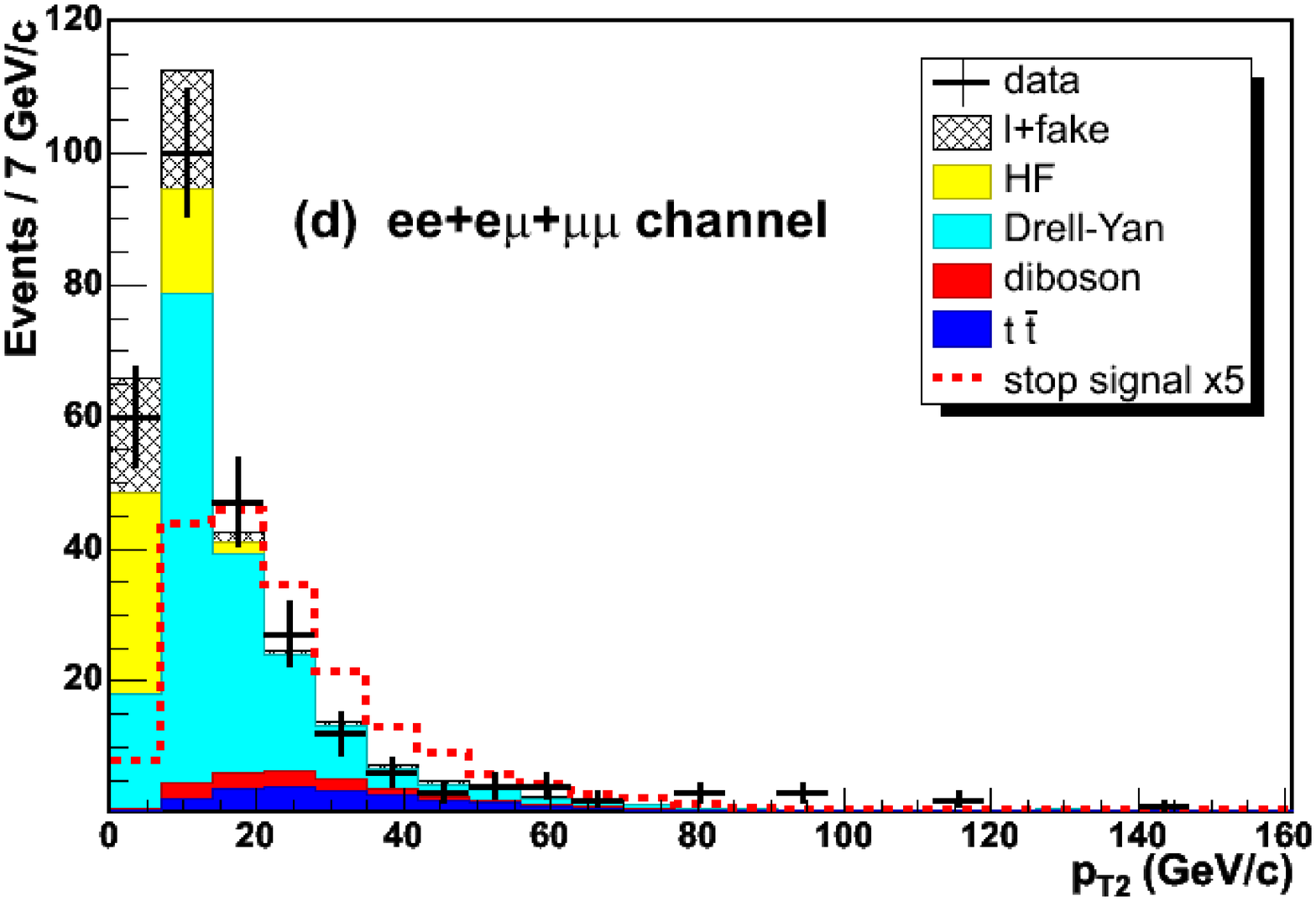}
 }
   \caption{\label{fig:PT2_presig}
$p_\mathrm{T}$ distributions of the next-to-highest $p_\mathrm{T}$ lepton in the pre-signal region,
shown as in Fig.~\ref{fig:MET_presig}. }
\end{figure*}

\begin{figure*}[htpb]
\centerline{\includegraphics[width=0.50\textwidth]{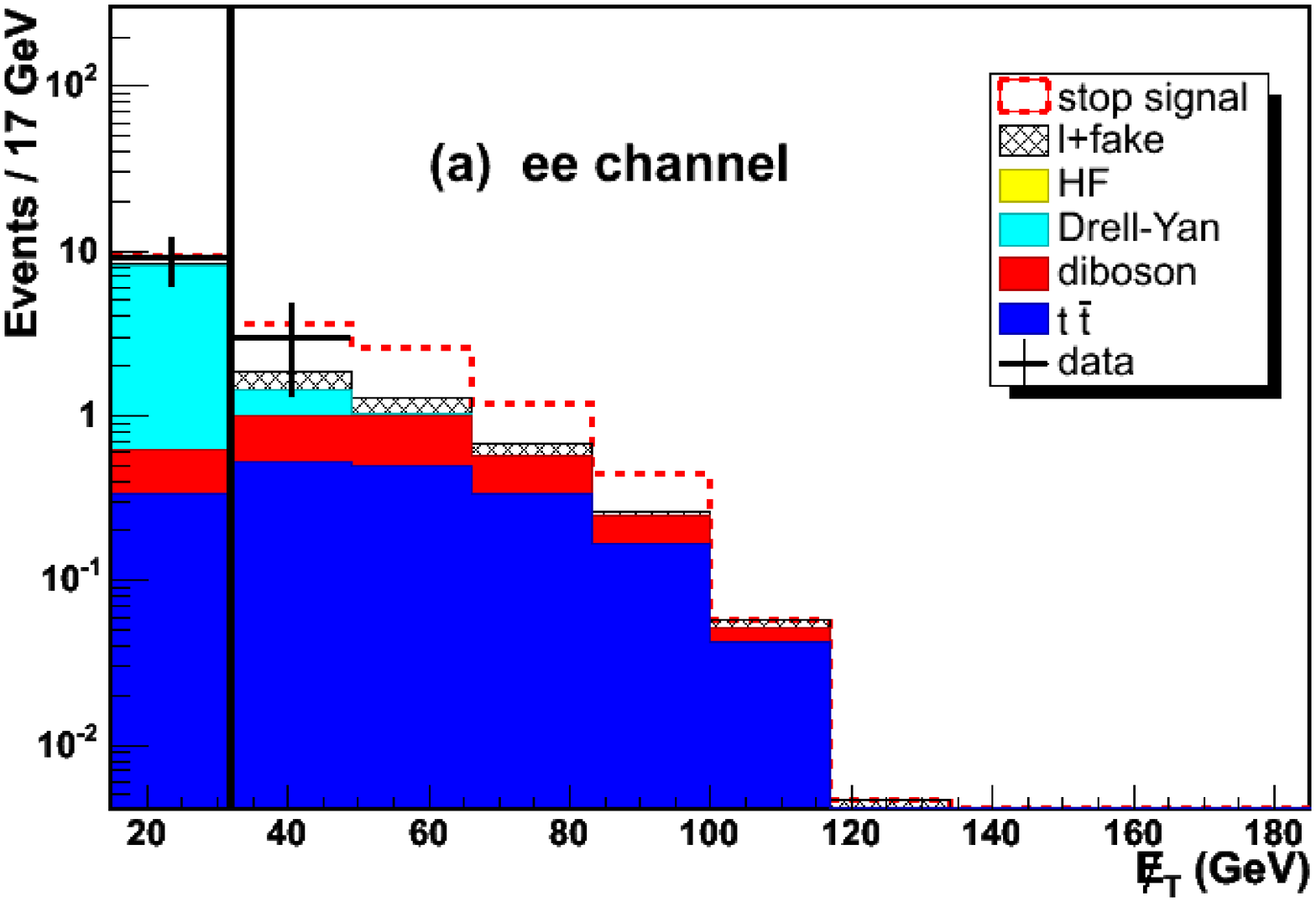}
   \hspace*{-.3cm}
   \includegraphics[width=0.50\textwidth]{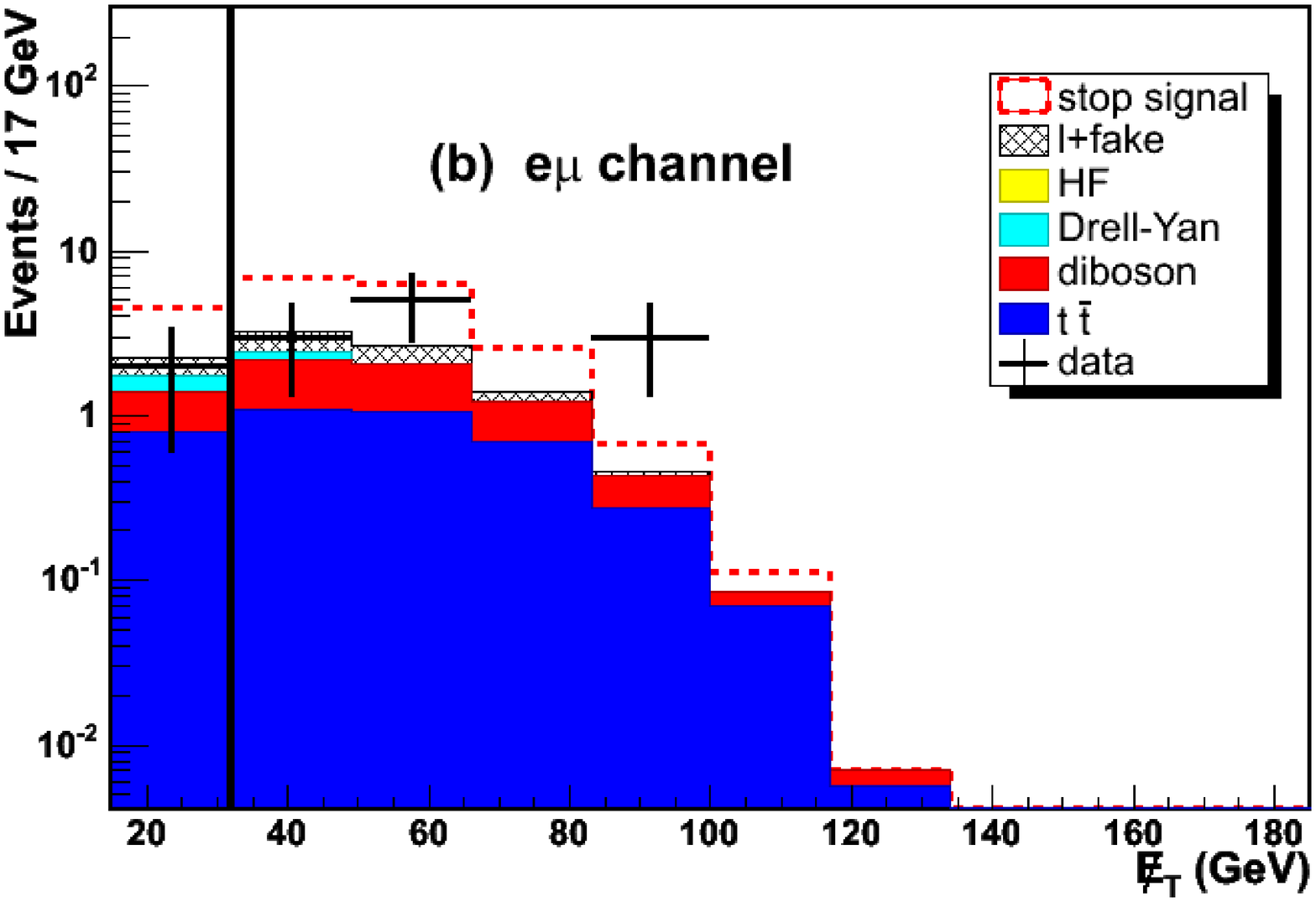}
 }
\centerline{\includegraphics[width=0.50\textwidth]{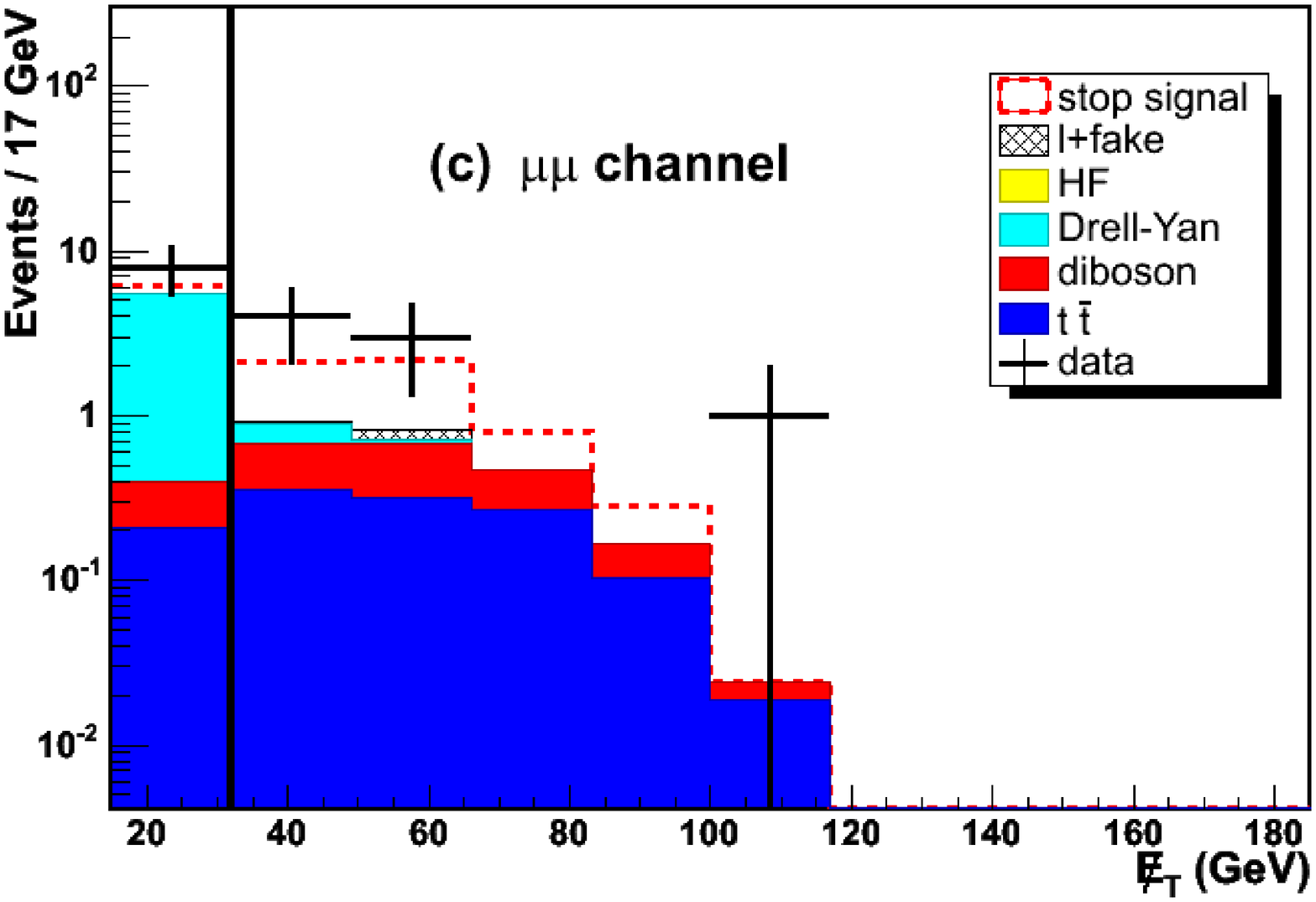}
   \hspace*{-.3cm}
   \includegraphics[width=0.50\textwidth]{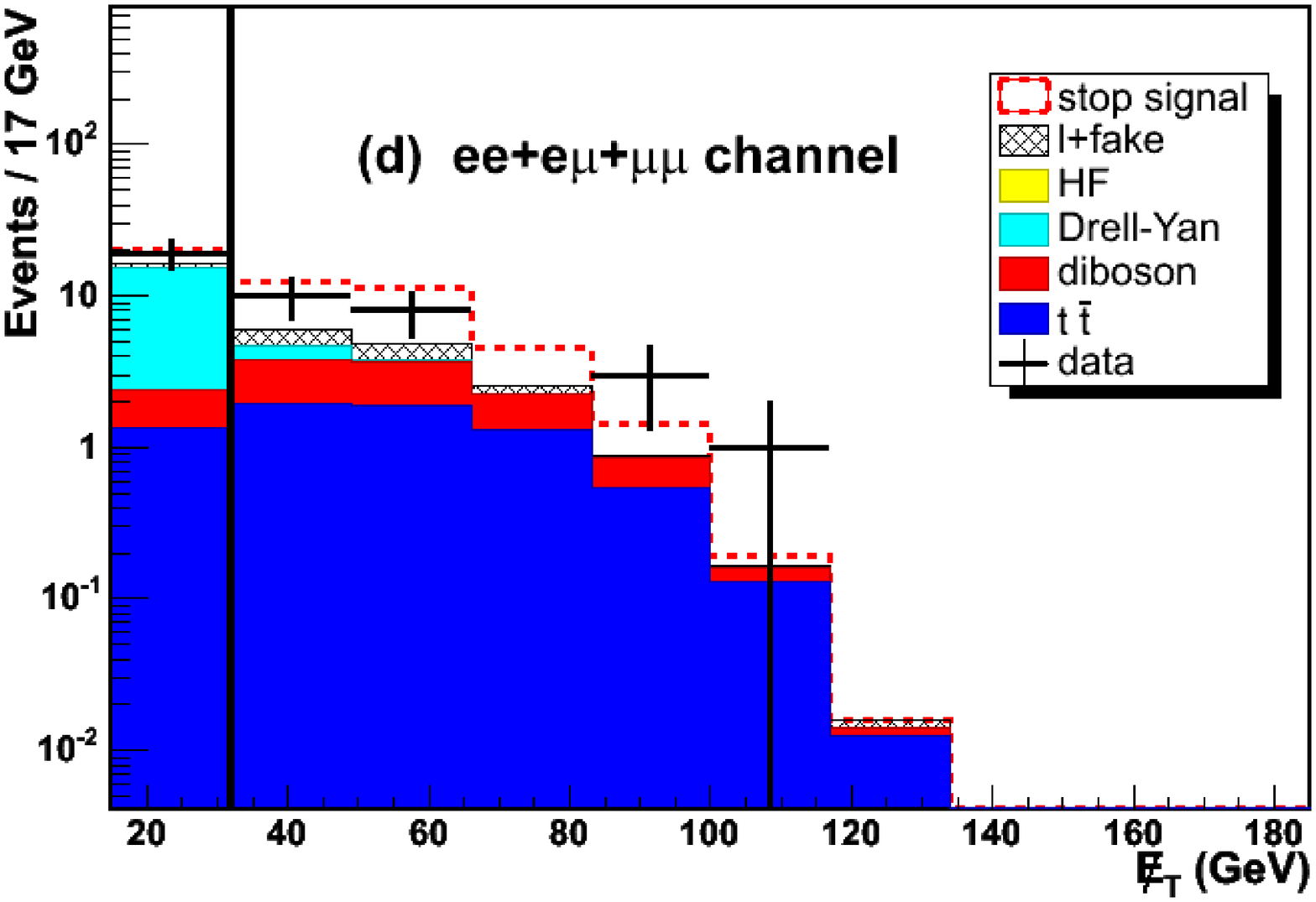}
 }
   \caption{\label{fig:MET_sigb}
The cut group $b$ $\not$$E_\mathrm{T}$ distributions for (a) the $ee$ channel, (b) the $e\mu$ channel,
(c) the $\mu\mu$ channel, and (d) the three channels combined with all other cuts
applied. Data are shown as the points with error bars (statistical only).
Shown as stacked histograms are the
backgrounds arising from misidentified hadrons and decays-in-flight ($l$+fake),
$b \bar b$ and $c \bar c$ (HF), DY, dibosons, and $t \bar t$.
MC predictions for $(m_{\tilde{t}},m_{\tilde{\nu}})$ = (140,90) GeV/$c^2$ are
stacked on top of the sums of backgrounds and are shown
as the dashed line. The cut selects events to the right of the vertical line.}
\end{figure*}

\begin{figure*}[htpb]
\centerline{\includegraphics[width=0.50\textwidth]{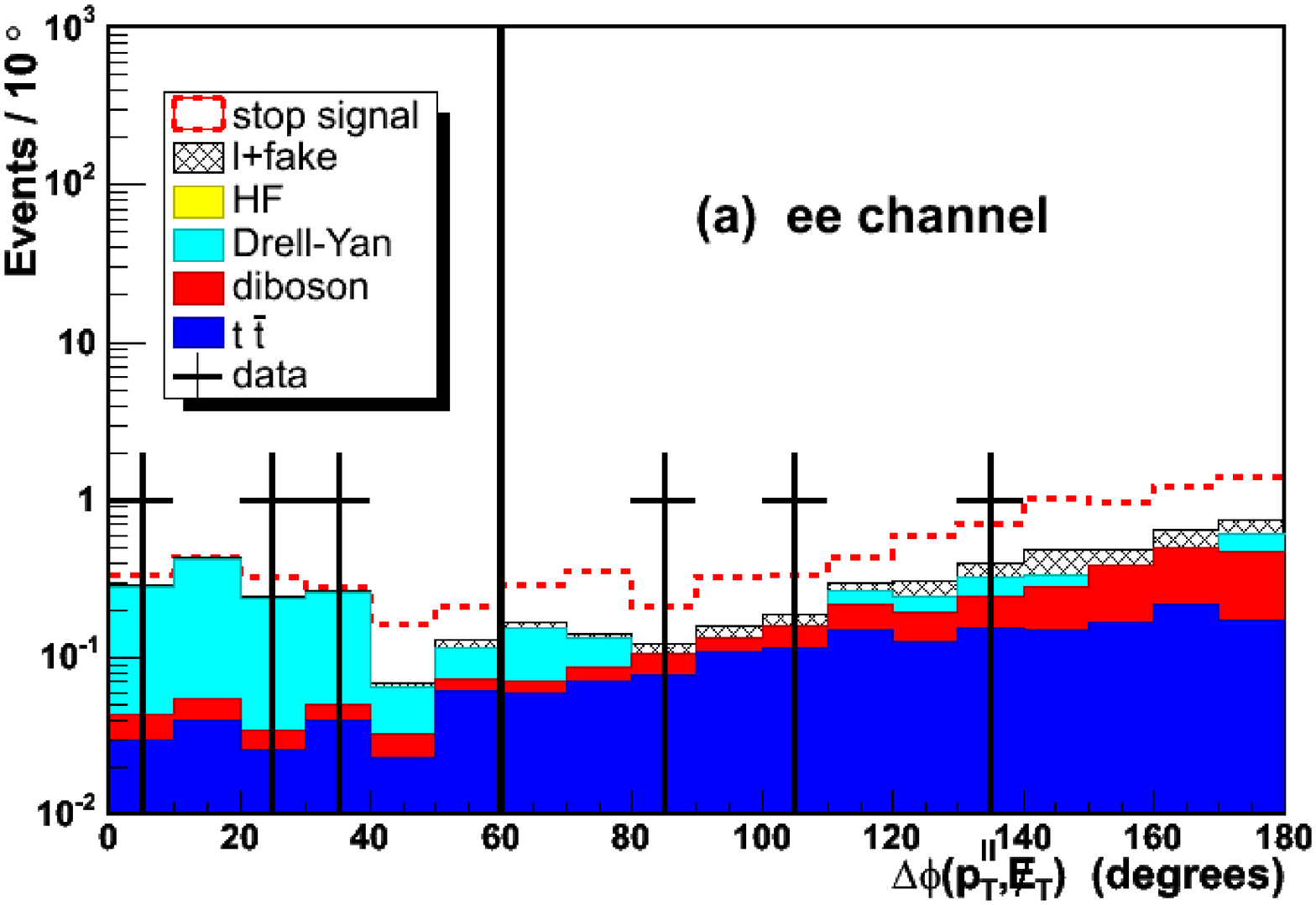}
   \hspace*{-.3cm}
   \includegraphics[width=0.50\textwidth]{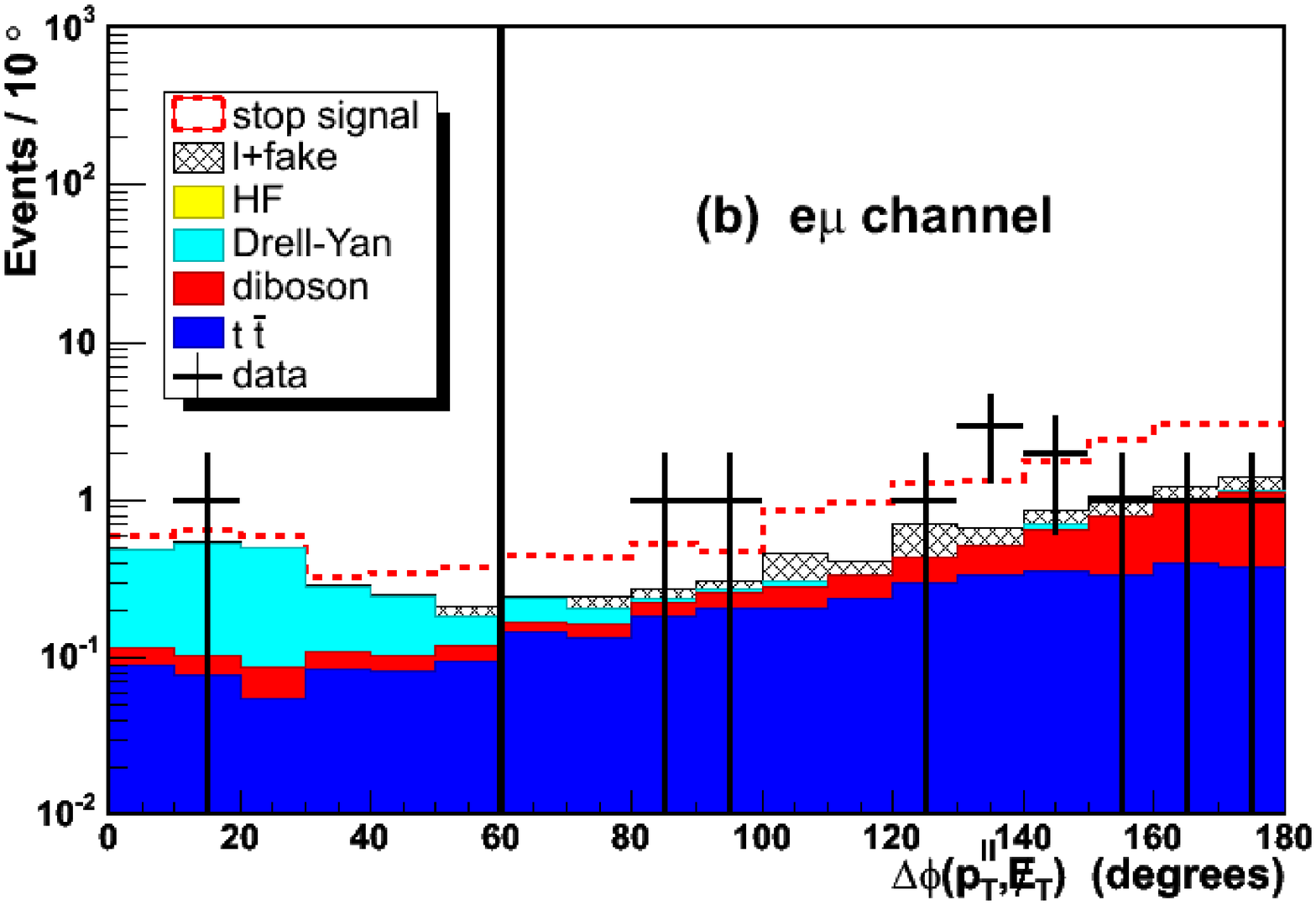}
 }
\centerline{\includegraphics[width=0.50\textwidth]{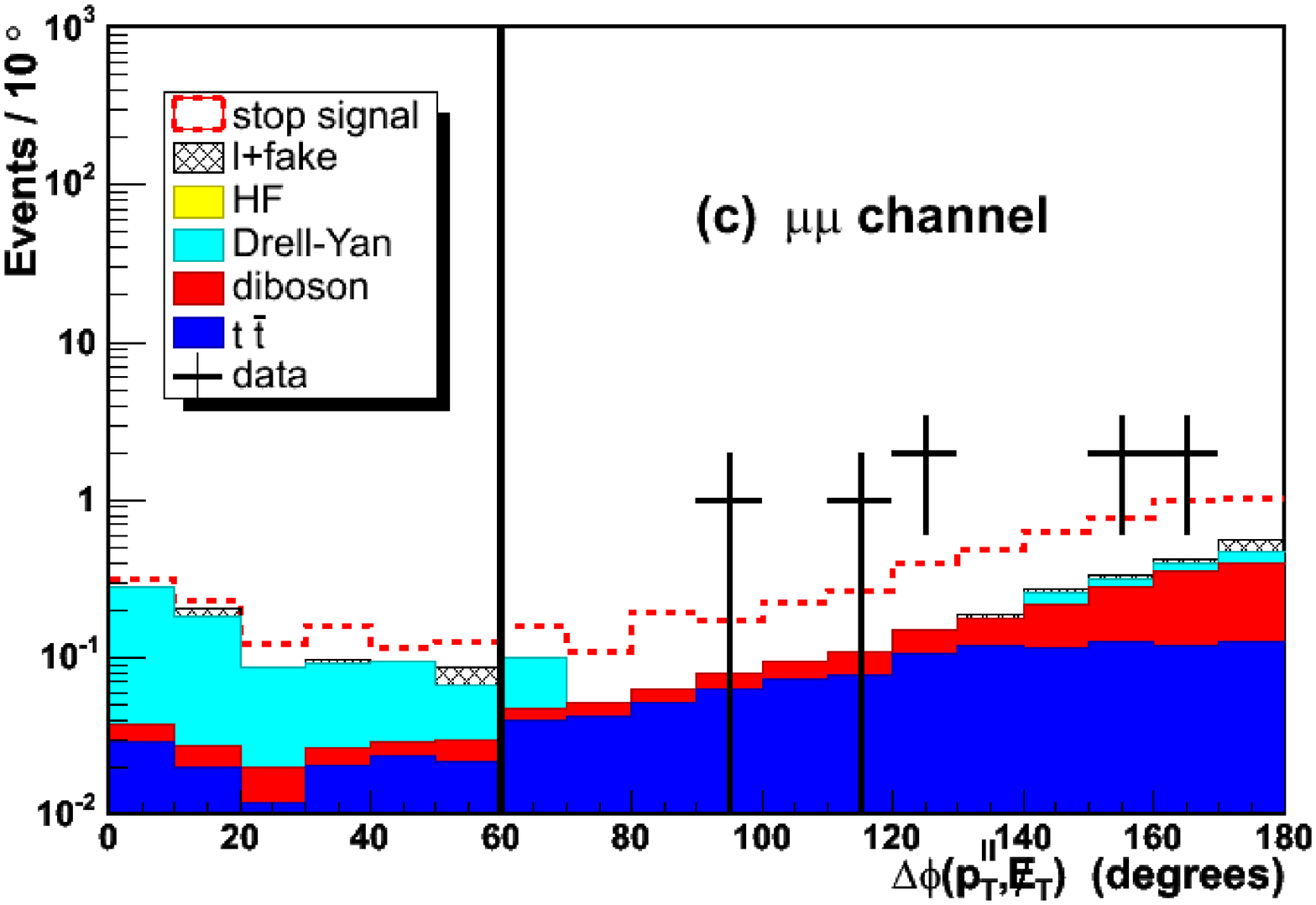}
   \hspace*{-.3cm}
   \includegraphics[width=0.50\textwidth]{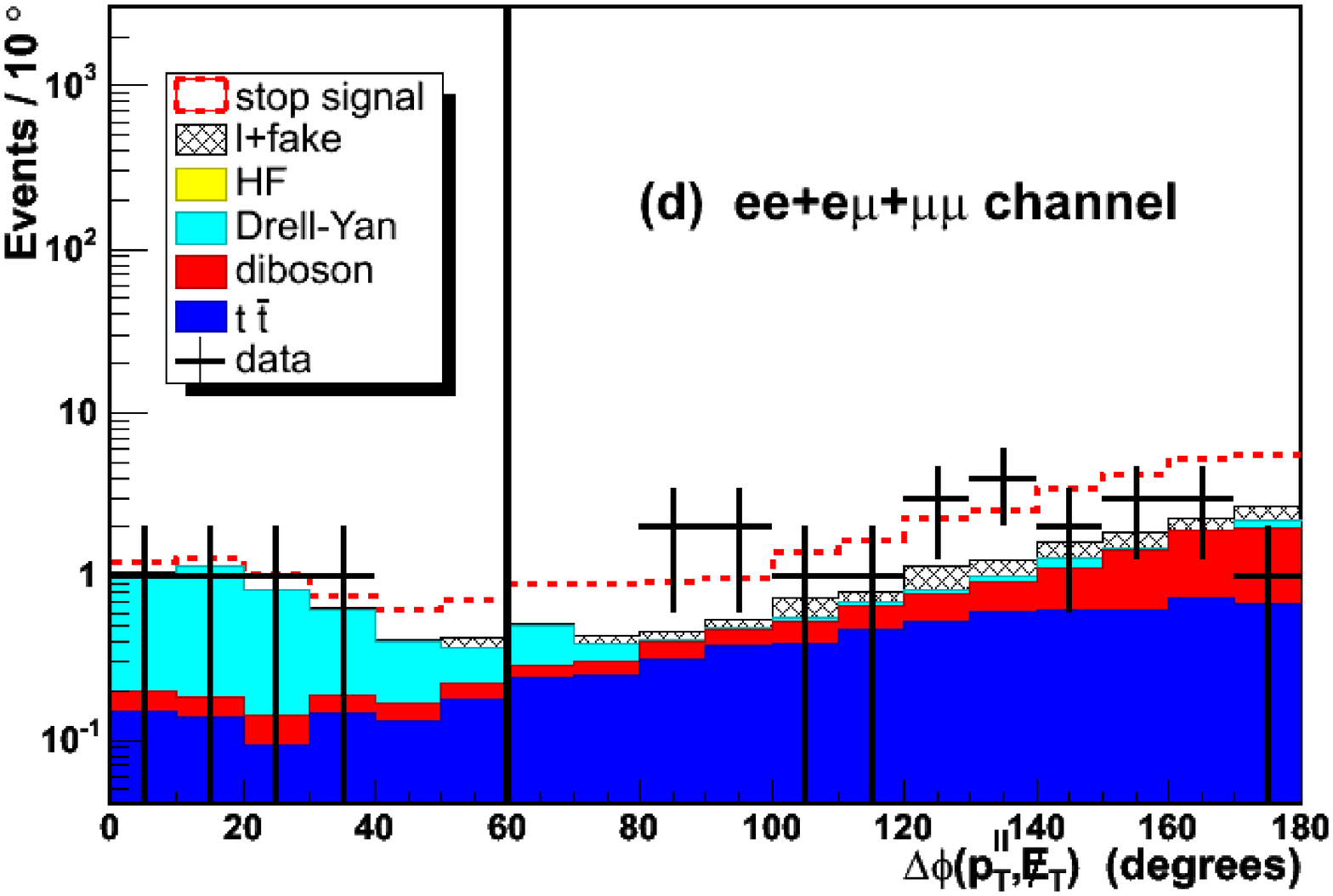}
 }
   \caption{\label{fig:phi12M_sigb}
The cut group $b$ distributions of the absolute value of the azimuthal angle between
the dilepton system and $\not$$E_\mathrm{T}$, shown as in Fig.~\ref{fig:MET_sigb}. }
\end{figure*}

\begin{figure*}[htpb]
\centerline{\includegraphics[width=0.50\textwidth]{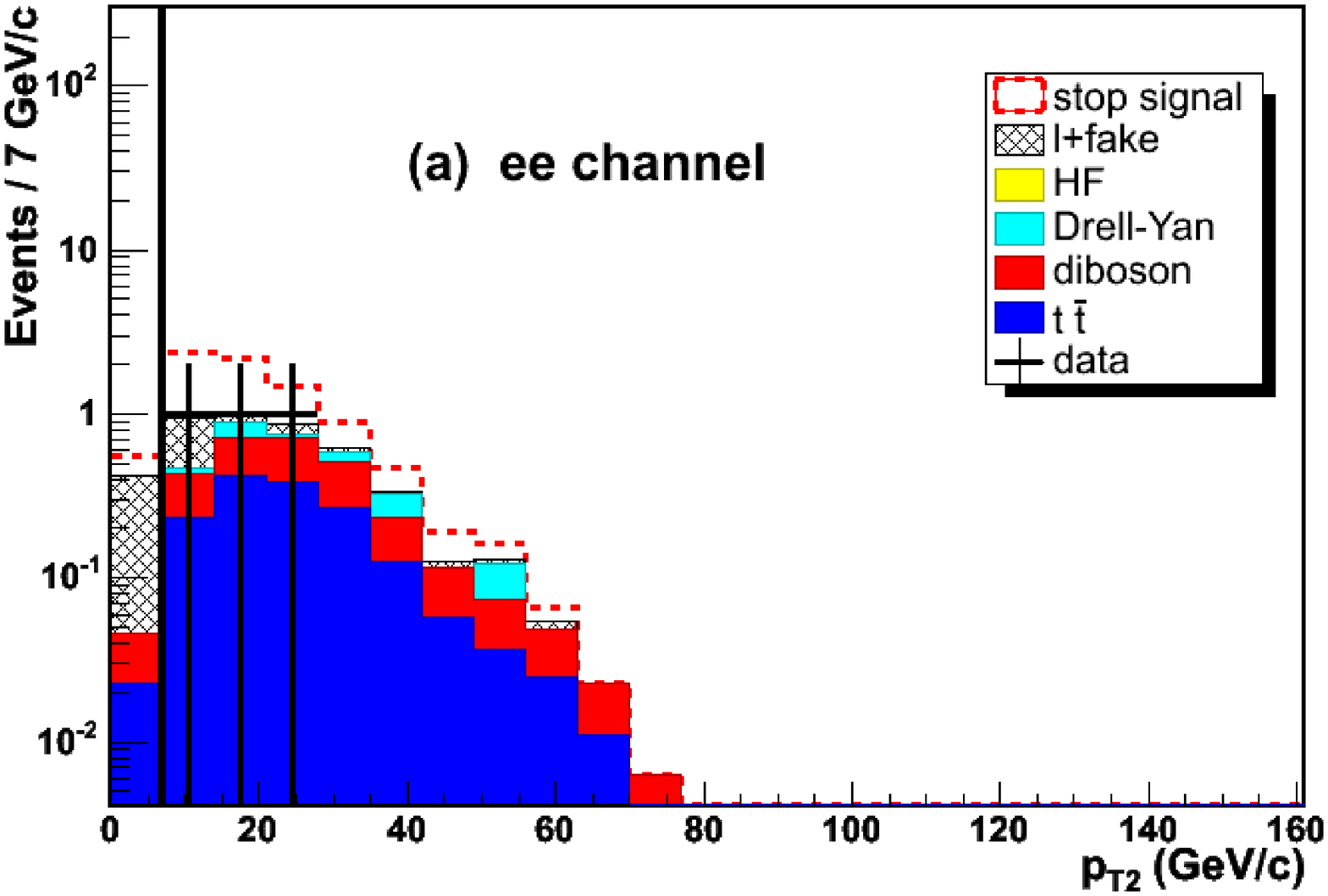}
   \hspace*{-.3cm}
   \includegraphics[width=0.50\textwidth]{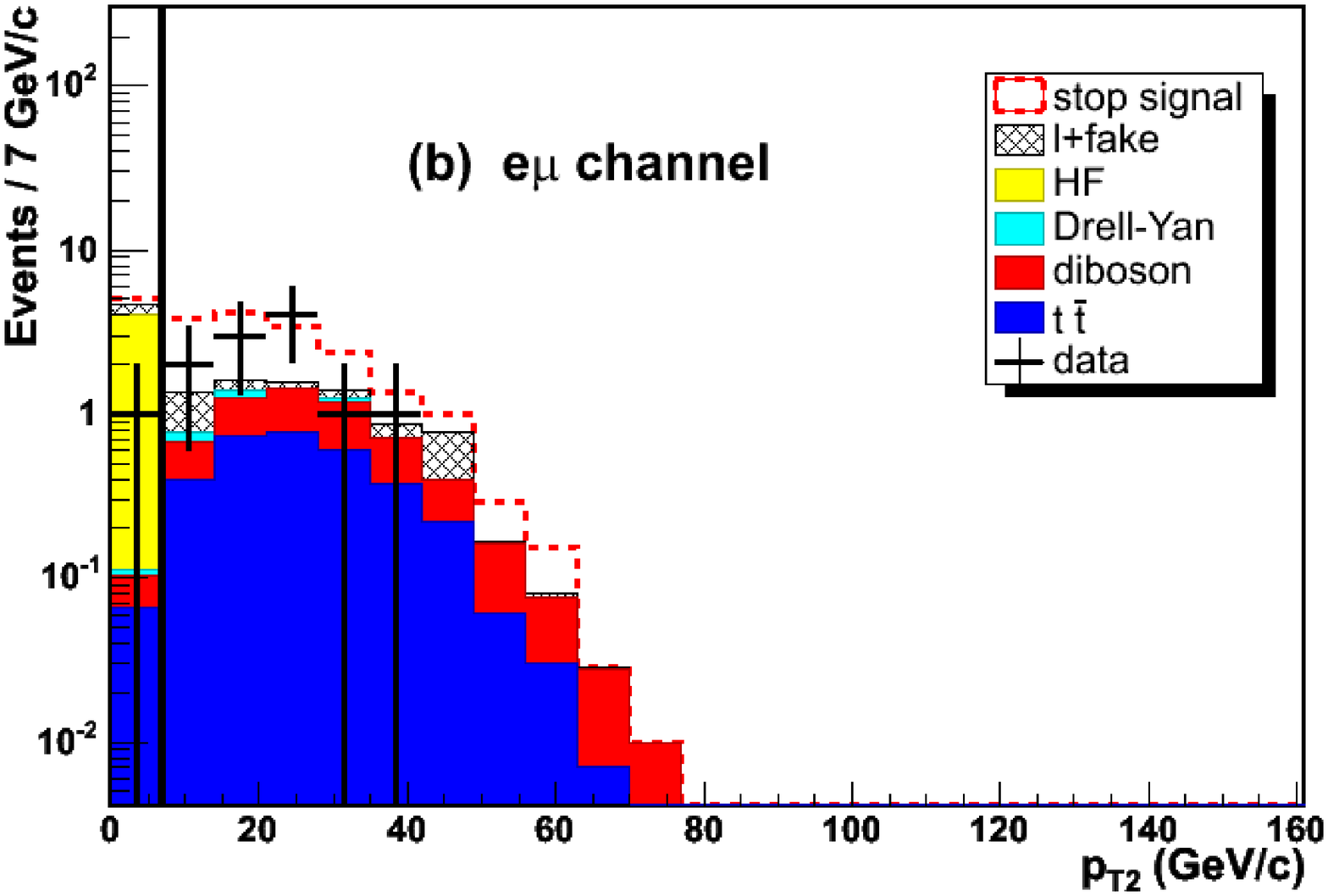}
 }
\centerline{\includegraphics[width=0.50\textwidth]{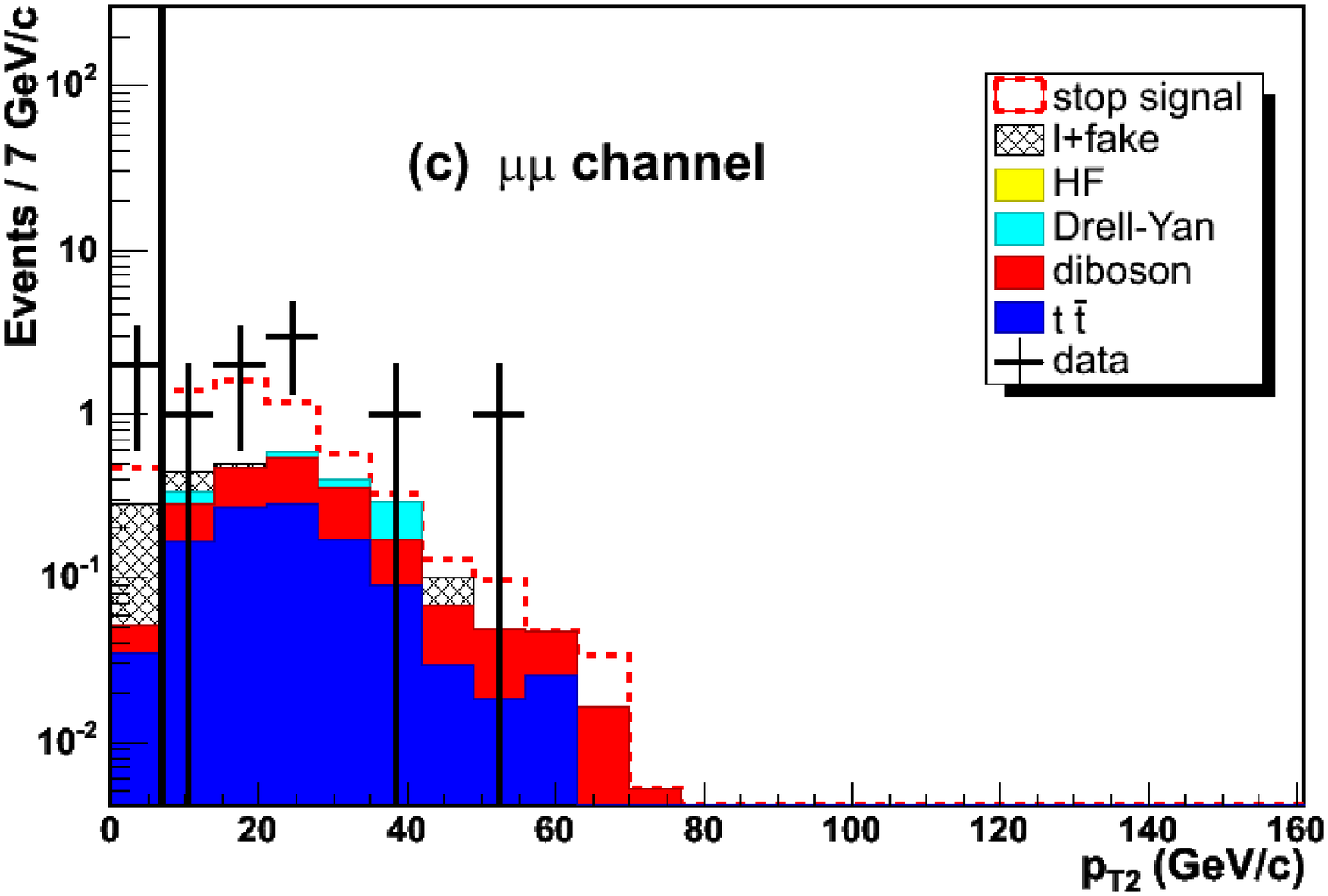}
   \hspace*{-.3cm}
   \includegraphics[width=0.50\textwidth]{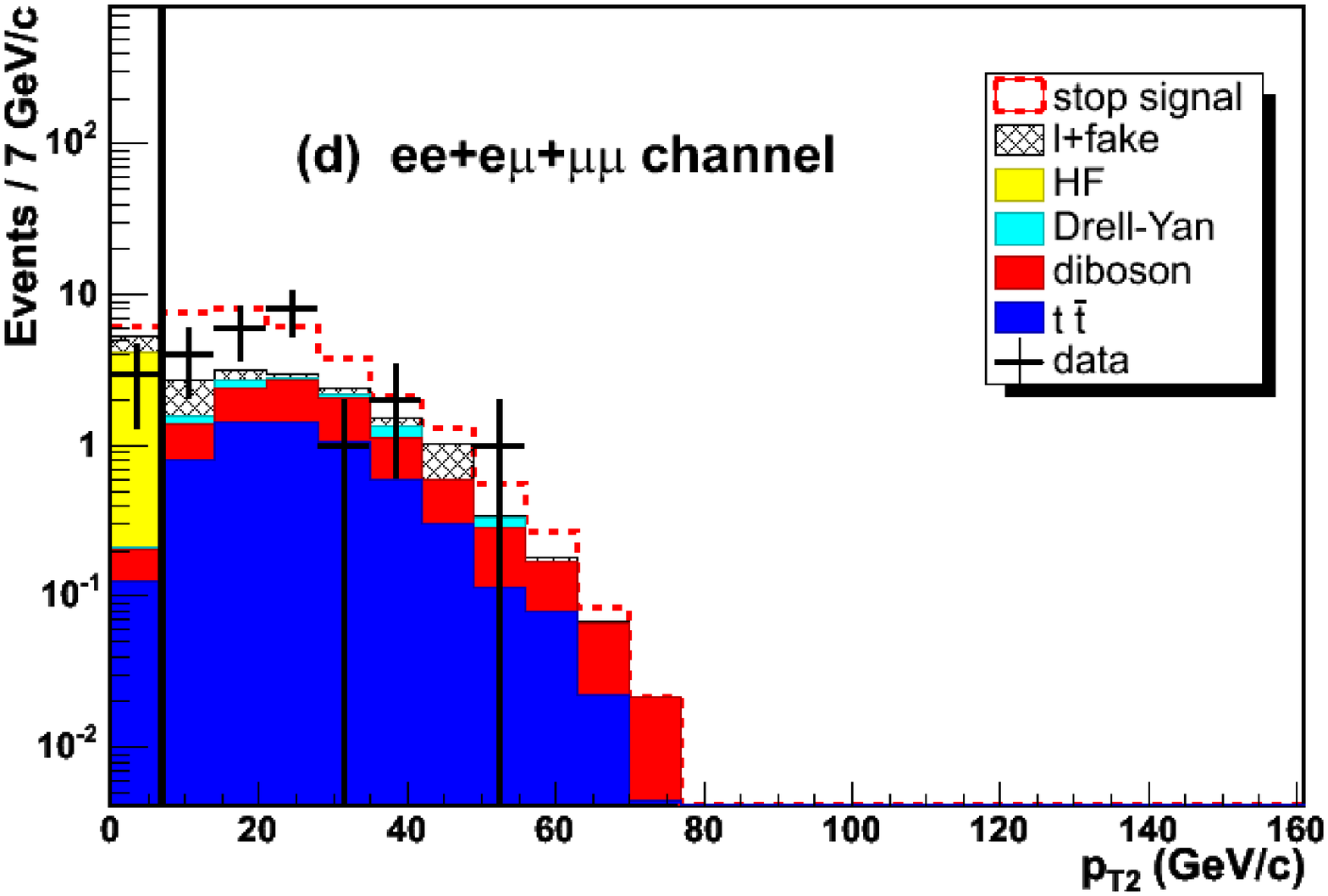}
 }
   \caption{\label{fig:pt2_sigb}
The cut group $b$ $p_\mathrm{T2}$ distributions, shown as in Fig.~\ref{fig:MET_sigb}. }
\end{figure*}

\begin{figure*}[htpb]
\centerline{\includegraphics[width=0.50\textwidth]{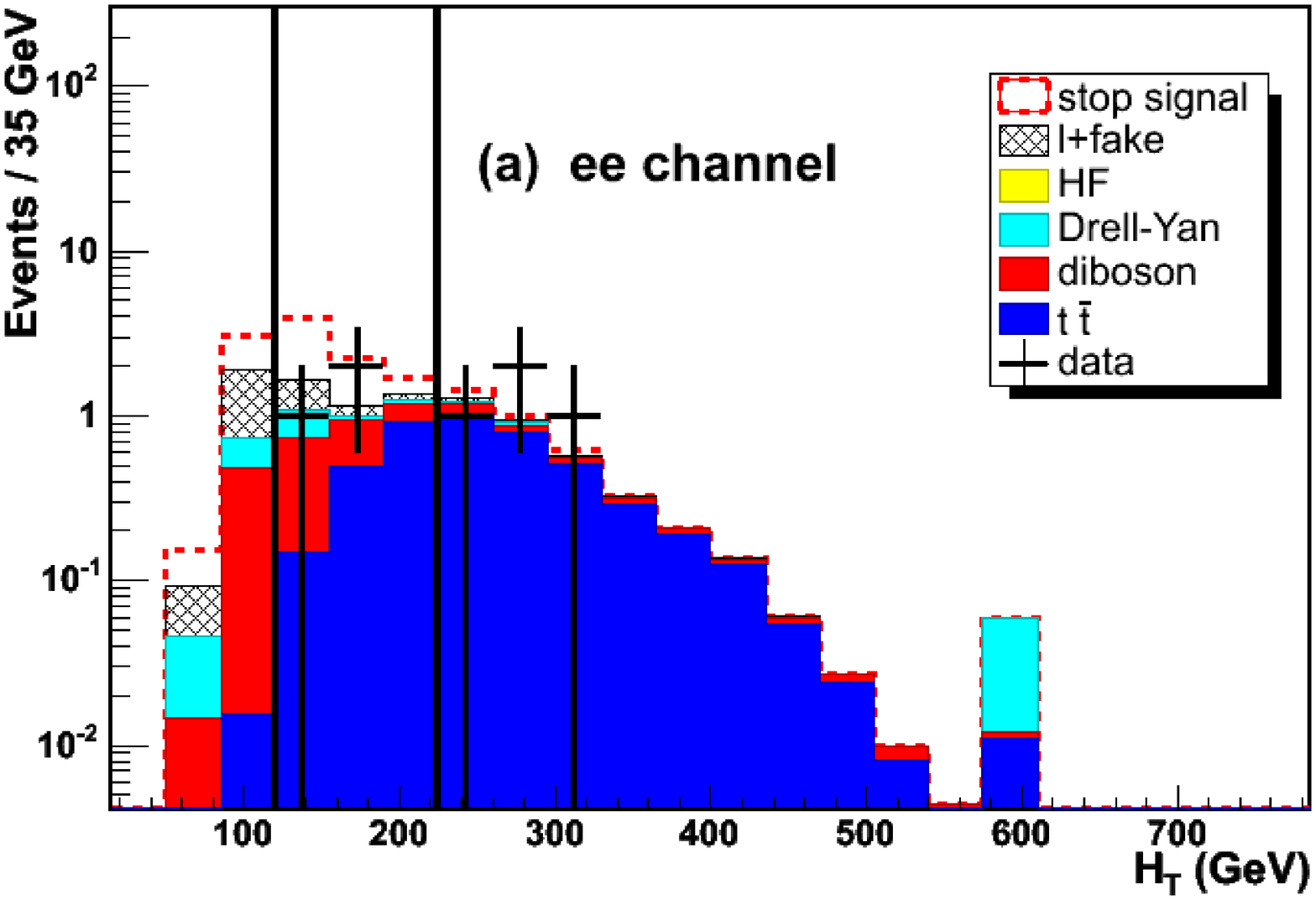}
   \hspace*{-.3cm}
   \includegraphics[width=0.50\textwidth]{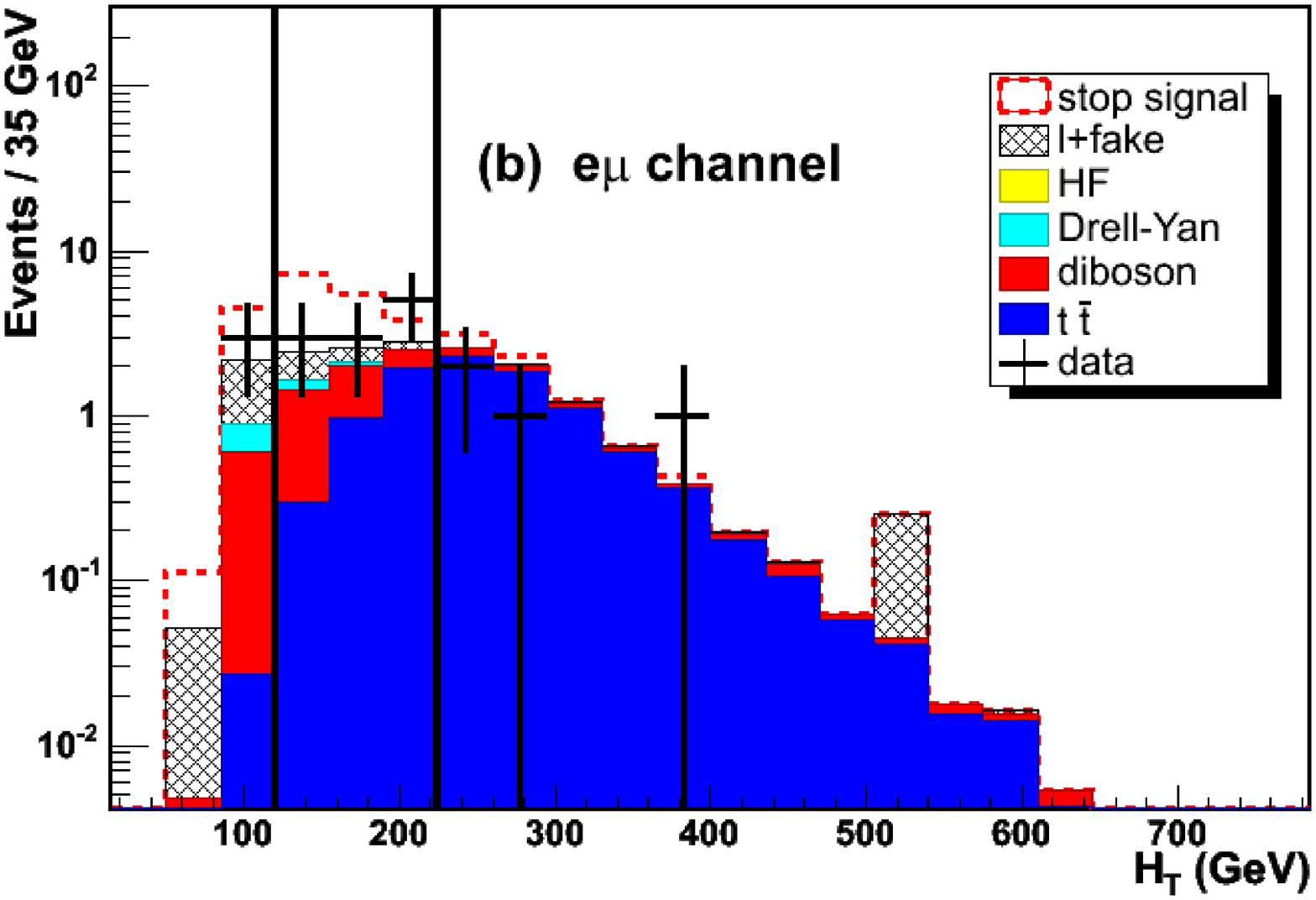}
 }
\centerline{\includegraphics[width=0.50\textwidth]{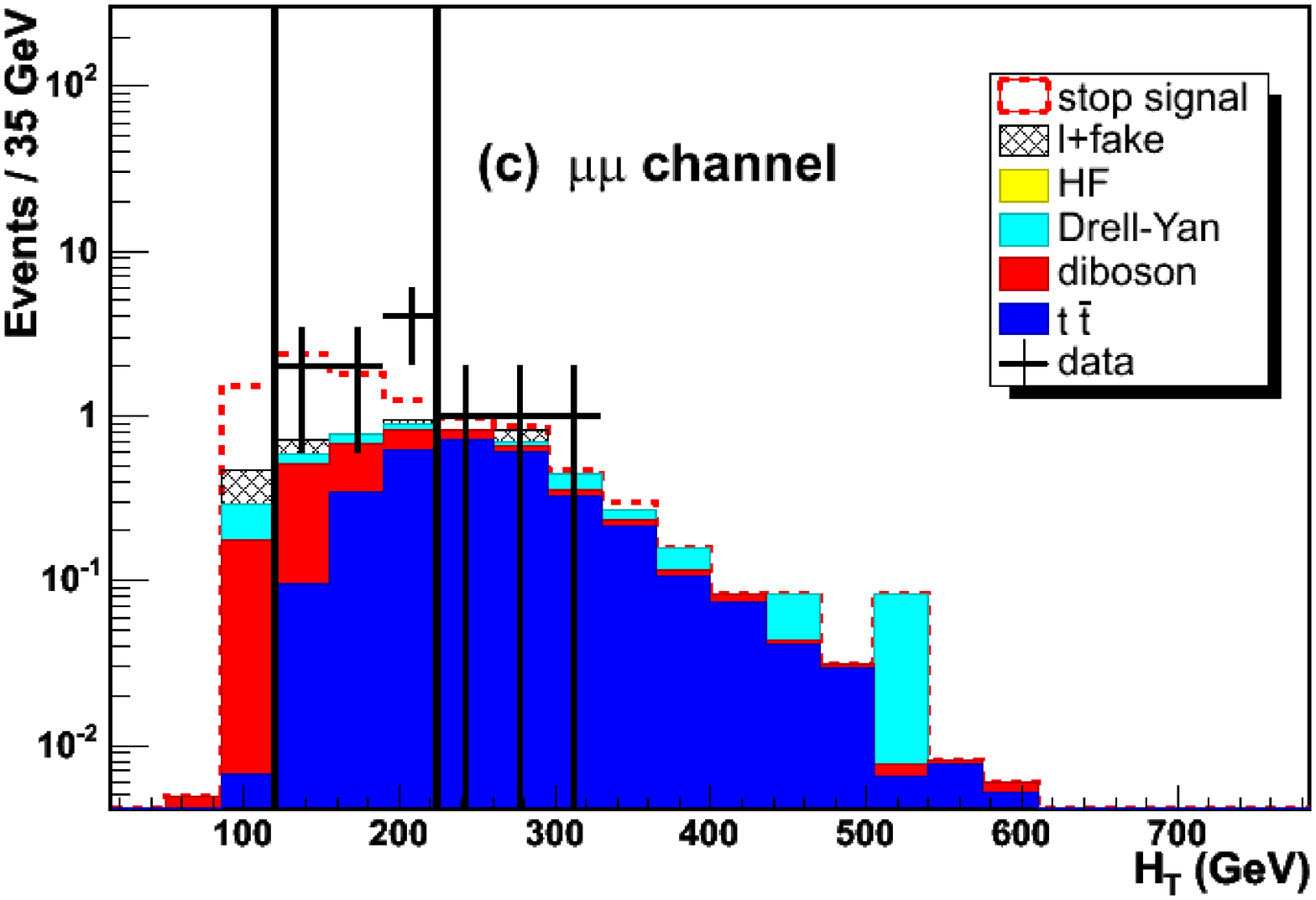}
   \hspace*{-.3cm}
   \includegraphics[width=0.50\textwidth]{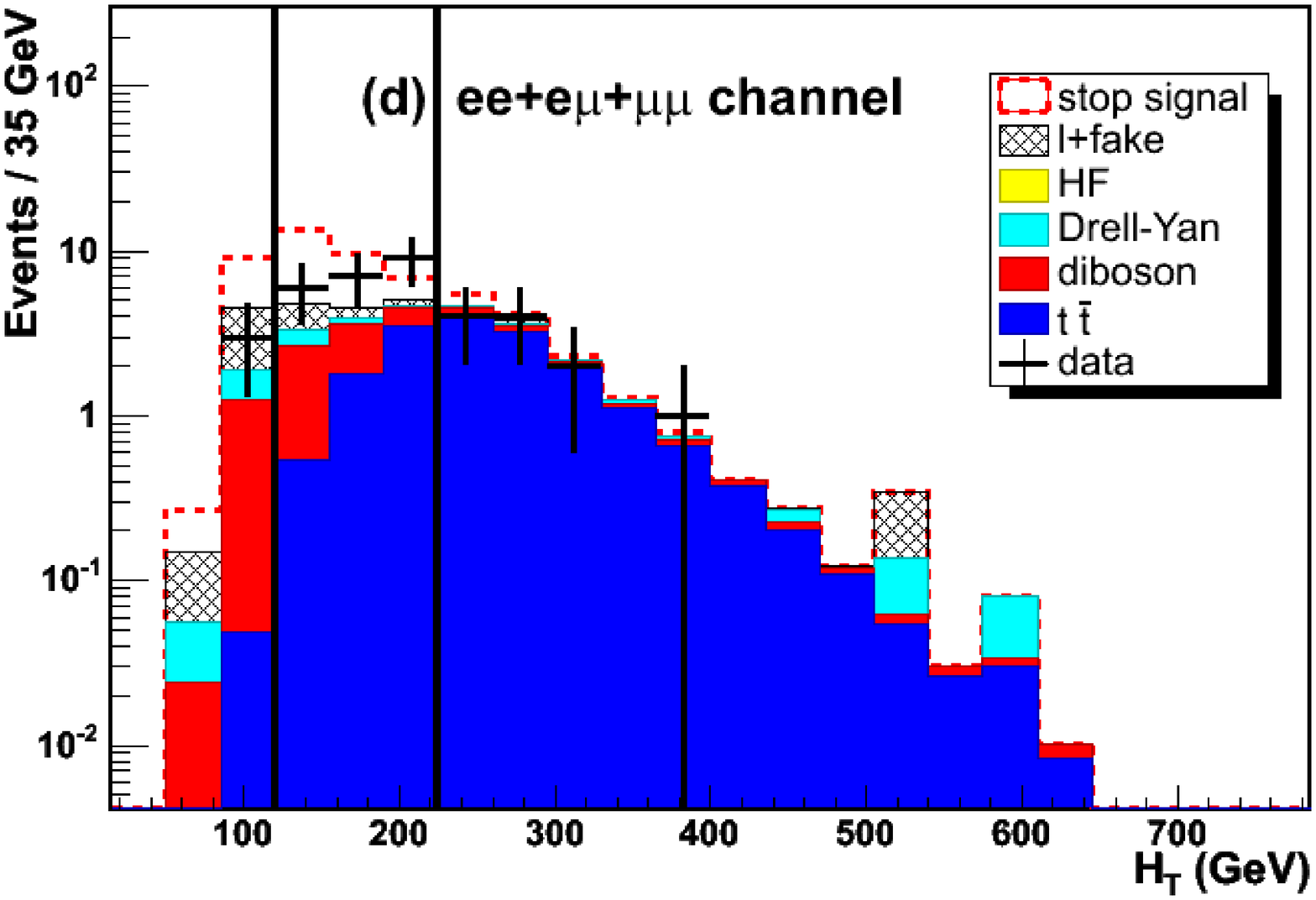}
 }
   \caption{\label{fig:Ht_sigb}
The cut group $b$ $H_\mathrm{T}$ distributions, shown as in Fig.~\ref{fig:MET_sigb}.
The cut selects events between the vertical lines. }
\end{figure*}

\begin{figure*}[htpb]
\centerline{\includegraphics[width=0.50\textwidth]{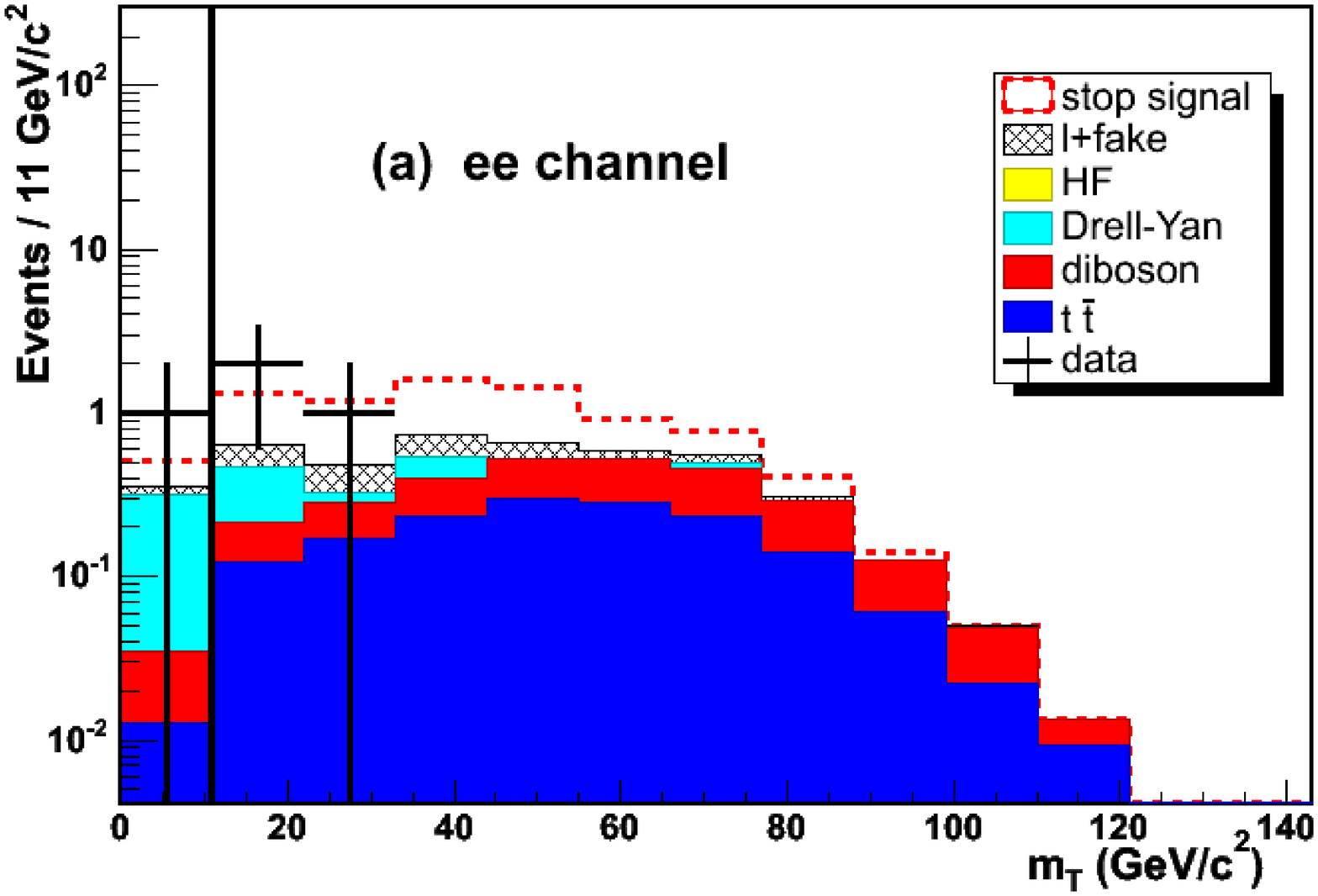}
   \hspace*{-.3cm}
   \includegraphics[width=0.50\textwidth]{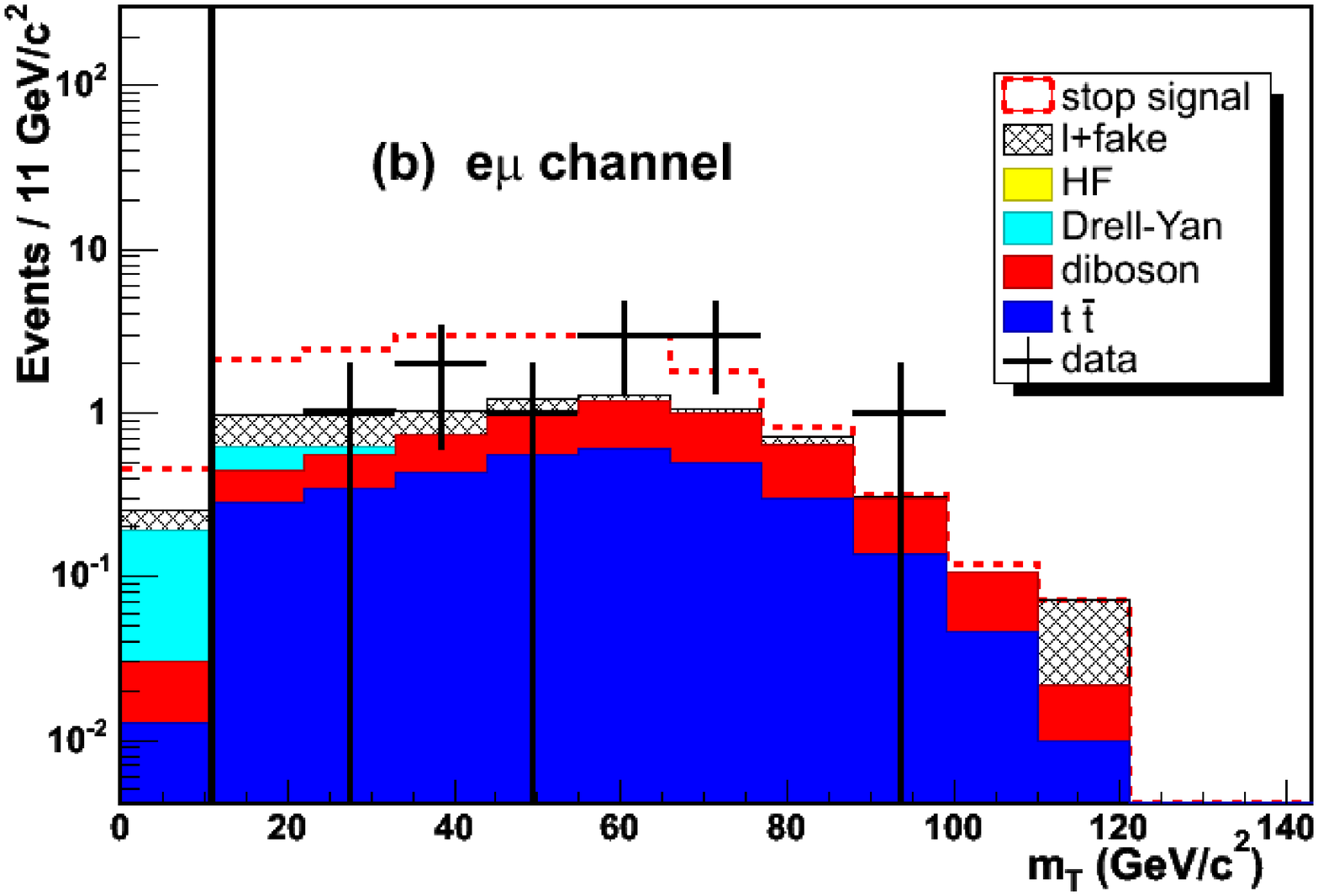}
 }
\centerline{\includegraphics[width=0.50\textwidth]{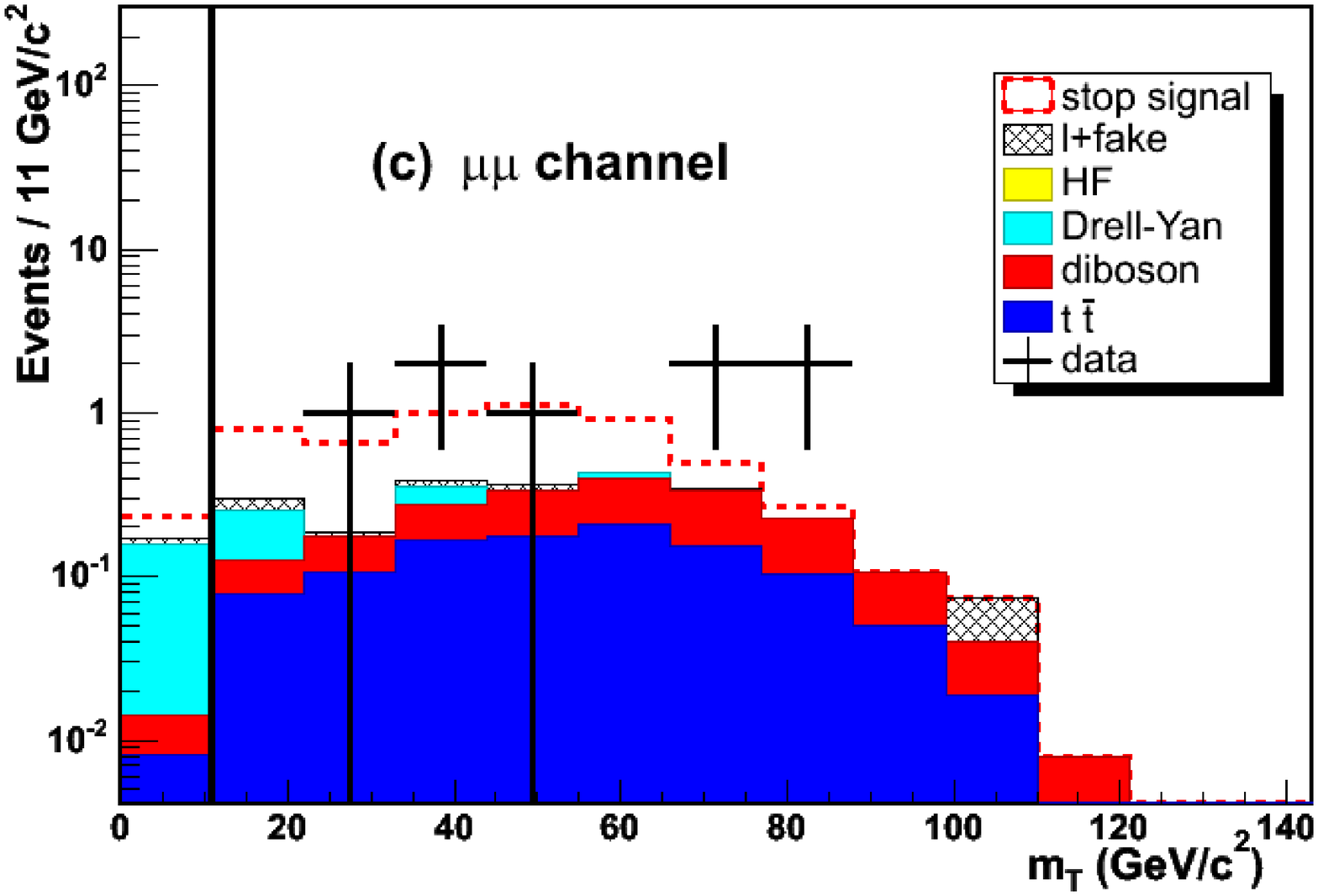}
   \hspace*{-.3cm}
   \includegraphics[width=0.50\textwidth]{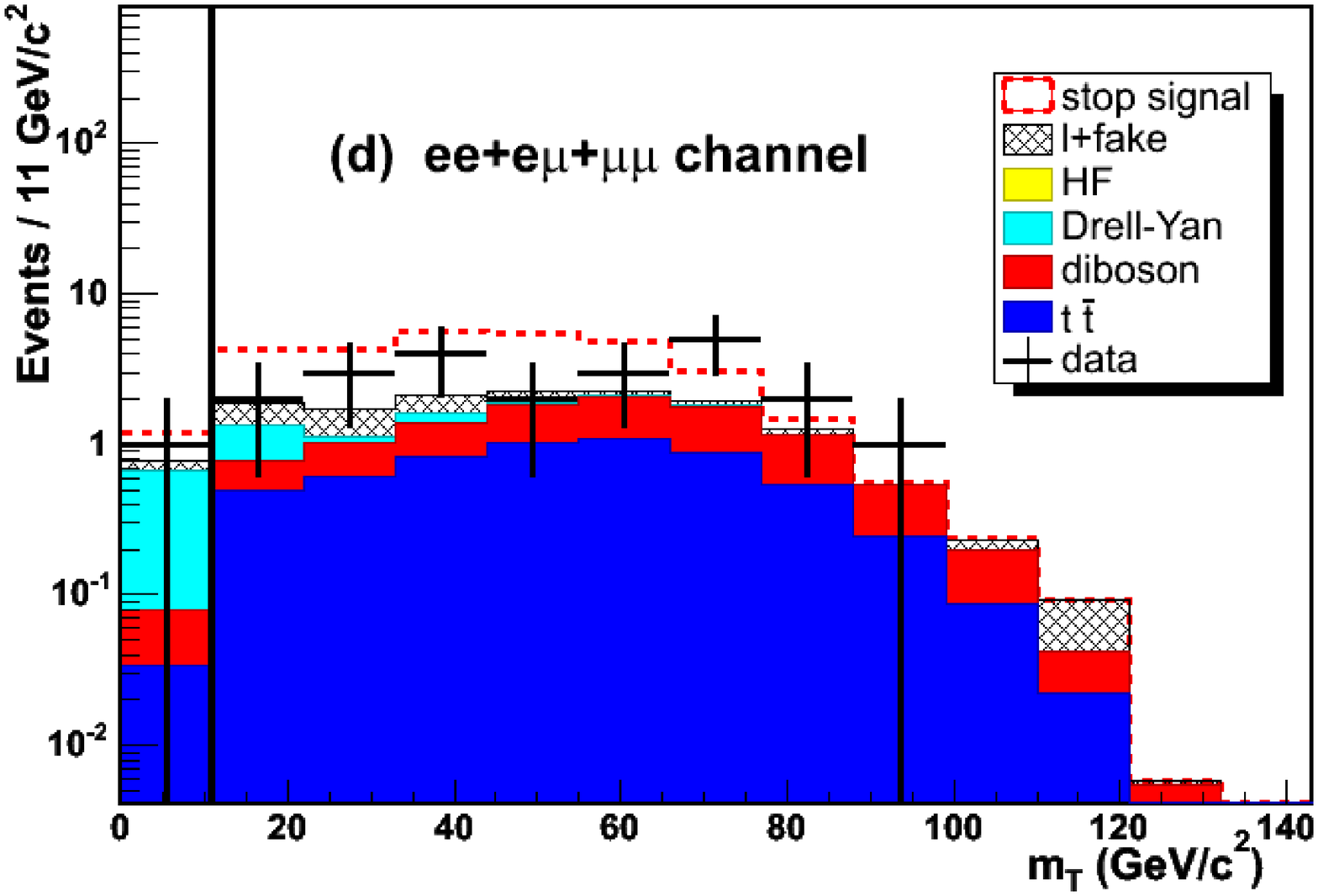}
 }
   \caption{\label{fig:Mt_sigb}
The cut group $b$ $m_\mathrm{T}$ distributions, shown as in Fig.~\ref{fig:MET_sigb}. }
\end{figure*}

\begin{figure*}[htpb]
\begin{center}
\includegraphics[width=0.8\textwidth]{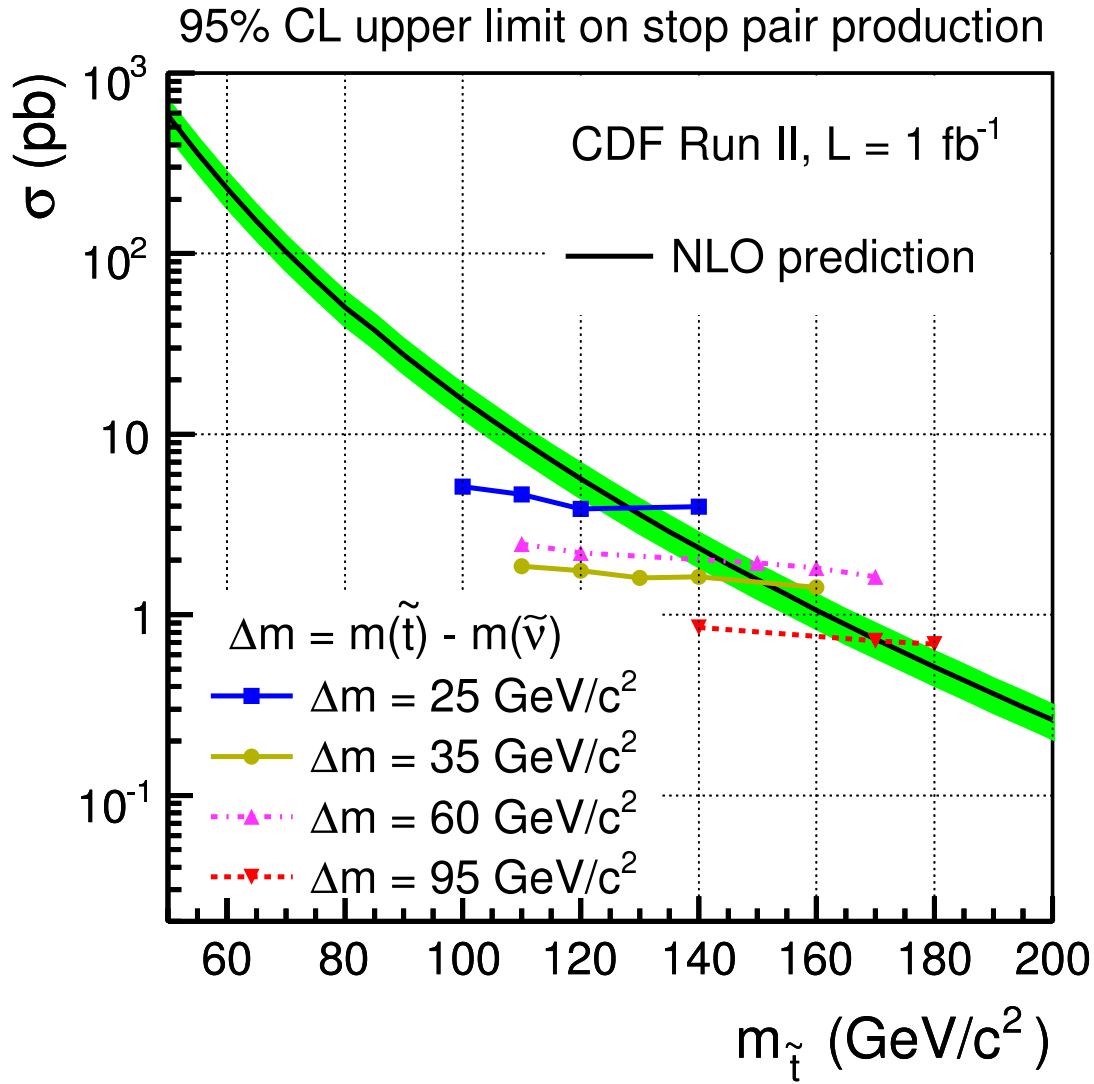}
\caption{\label{fig:xs2} Stop cross section upper limits for fixed stop-sneutrino mass differences.
The NLO {\sc prospino} cross section with CTEQ6M PDFs is shown as the solid curve.
The band represents
the theoretical uncertainty in the cross section due to uncertainties on the renormalization and
factorization scales and the PDFs.}
\end{center}
\end{figure*}

\begin{figure*}[htpb]
\begin{center}
\includegraphics[width=0.8\textwidth]{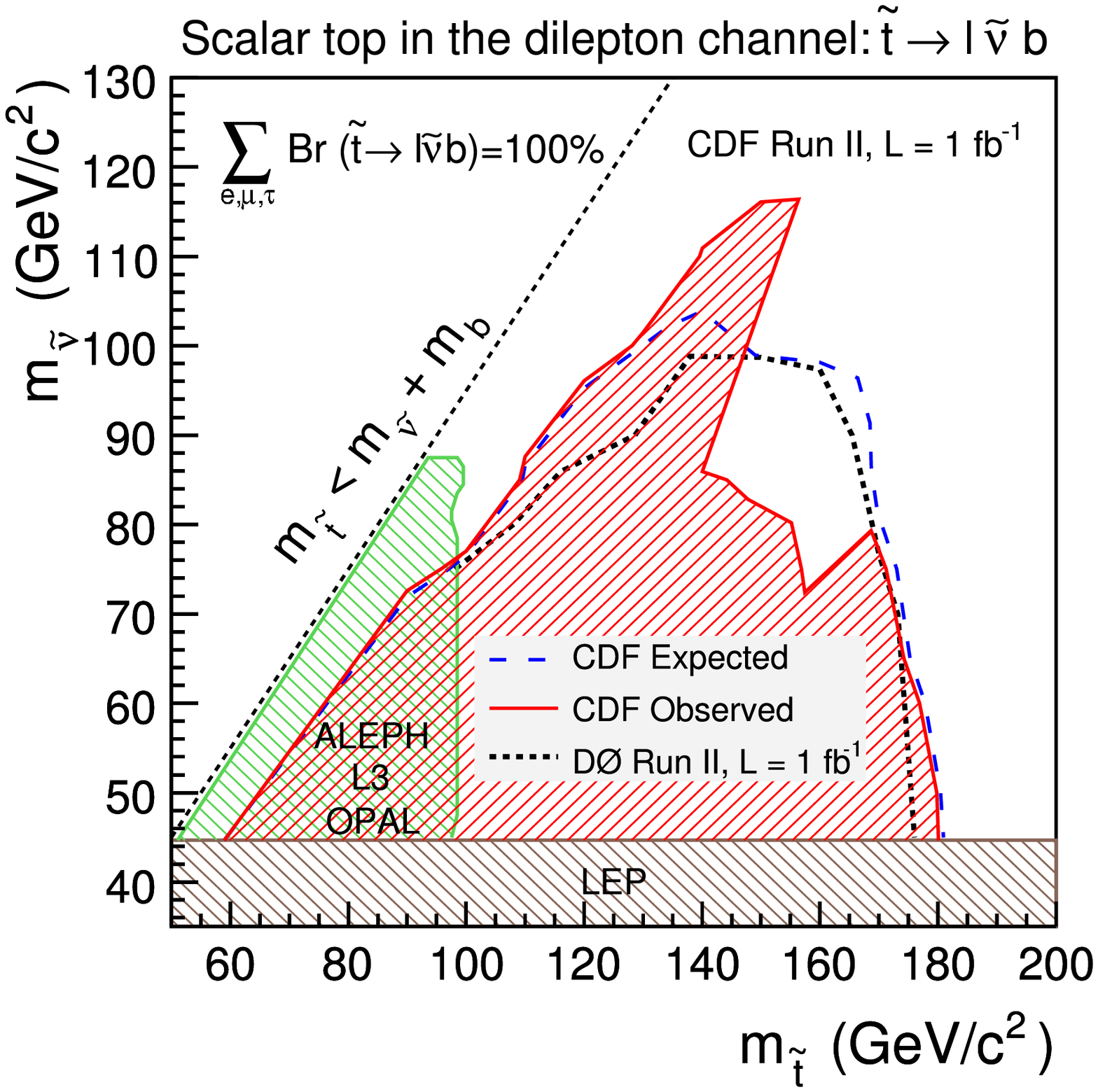}
\caption{\label{fig:obsL} Observed and expected limits in the stop-sneutrino mass plane.
The LEP limits are for a mixing angle of zero which provides the greatest reach.}
\end{center}
\end{figure*}

\section{THE DETECTOR AND DATA SET}

The data sample corresponds to an integrated luminosity of
1 fb$^{-1}$ of \ppbar~collisions at a
center-of-mass energy of 1.96~TeV collected
with the CDF detector \cite{CDF} at the Fermilab Tevatron.
Of particular relevance to this analysis are the tracking system,
the calorimetry, and the muon detectors.

The tracking system consists of two silicon micro-strip detectors
and a multi-wire open-cell drift chamber, the central outer tracker (COT).
The silicon vertex detector (SVX) and the intermediate silicon layers
cover the pseudorapidity \cite{met} region  $|\eta|<2$ while the COT covers $|\eta|<1$.
The tracking system is surrounded by a superconducting solenoid
with a magnetic field of 1.4 T. The relative track momentum resolution provided by
the COT is $\sigma_{p_\mathrm{T}}/{p_\mathrm{T}}^2\simeq0.0015${(GeV/$c$)}$^{-1}$.

Outside the magnet in the radial direction are electromagnetic
and hadronic calorimeters arranged in a projective tower geometry
with a tower granularity of $\Delta\eta \times \Delta\phi\simeq 0.1 \times 0.26$ in the
central region.
The central electromagnetic calorimeter (CEM) utilizes lead-scintillator
sampling and covers $|\eta|<1.1$, with energy resolution for electrons
$\sigma_{E_\mathrm{T}}/E_\mathrm{T}\simeq13.5\%/\sqrt{E_\mathrm{T}~{\rm (GeV)}}\oplus2\%$.
The central electromagnetic shower maximum detector (CES)
is located inside the CEM at a depth of six radiation lengths, close
to the position of maximum electromagnetic
shower development, and is used for the determination of the shower shape and for
matching the shower location to the track extrapolation.
The central hadron calorimeter (CHA) uses iron-scintillator
sampling and covers $|\eta|<0.9$. Its energy resolution for hadrons is
$\sigma_{E_\mathrm{T}}/E_\mathrm{T}\simeq75\%/\sqrt{E_\mathrm{T}~{\rm (GeV)}}\oplus3\%$.

Additional calorimetry
extends the coverage in the forward direction and is used in
this analysis for calculating
$\not$$E_\mathrm{T}$ and jet energies
but not for lepton identification. The plug electromagnetic calorimeter
covers $1.1<|\eta|<3.6$ and is constructed of lead and scintillator layers with an
energy resolution for electrons of
$\sigma_{E_\mathrm{T}}/E_\mathrm{T}\simeq16\%/\sqrt{E_\mathrm{T}~{\rm (GeV)}}\oplus1\%$.
The plug hadron calorimeter
covers $1.3<|\eta|<3.6$ and is constructed of iron and scintillator layers with an
energy resolution for hadrons of
$\sigma_{E_\mathrm{T}}/E_\mathrm{T}\simeq80\%/\sqrt{E_\mathrm{T}~{\rm (GeV)}}\oplus5\%$.
The iron and scintillator
wall hadron calorimeter covers the intermediate region  $0.7<|\eta|<1.3$
between the central and plug hadron calorimeters. Its energy resolution is
$\sigma_{E_\mathrm{T}}/E_\mathrm{T}\simeq75\%/\sqrt{E_\mathrm{T}~{\rm (GeV)}}\oplus4\%$.

Radially outside the calorimetry is the
muon detection system. The parts of the muon detector relevant
to this analysis are the central muon detector (CMU), the central muon
upgrade detector (CMP) and the central muon extension (CMX). The CMU consists
of four-layer drift chambers and covers the pseudorapidity
range  $|\eta|<0.6$. The CMP is made of four layers of single-wire
drift cells located
behind an additional 3.3 interaction lengths of steel and covers
$|\eta|<0.6$. Muons with reconstructed track stubs found in both the CMU and CMP
are labeled CMUP muons. The CMX extends the coverage to $|\eta|<1$ and is made
up of eight layers of drift tubes.

The data were collected using a three-level trigger system.
The first two levels are mostly hardware-based while the third level is
software-based and is a fast version of the offline event reconstruction package.
The online selection requires at least
two lepton candidates falling into the following categories:
CEM-CEM, CEM-CMUP, CEM-CMX, CMUP-CMUP, and CMUP-CMX, where the
leptons are labeled by the detector components used in their
identification. The triggers used in this analysis had an
$E_\mathrm{T}$ threshold of 4 GeV for electrons and a $p_\mathrm{T}$ threshold of 4 GeV/$c$ for muons.

All lepton candidates require the presence of a well-reconstructed track in the COT detector.
Offline, electrons are required to pass a $\chi^{2}$ comparison of the CES lateral shower
profile in the $r-z$ view  and the profile extracted from electron test-beam data.
The shower position in the CES must also match the extrapolated position of the track.
In addition, the lateral shower profile
in the CEM towers must be consistent with that expected from test-beam data.
The energy deposited in the
calorimeters must be consistent with the track momentum measurement and the energy deposition
in the CHA must be small. To reduce the background arising from the decay of hadrons in jets,
the electrons are required to be isolated.
Isolated electrons are selected by requiring that the remaining
transverse energy after subtracting the transverse energy associated with the electron
in a cone of 0.4 in $\eta-\phi$ space ($E_\mathrm{T}^{iso}$) must be
less than 2 GeV if $E_\mathrm{T}\le 20$ GeV
or $E_\mathrm{T}^\mathrm{iso}/E_\mathrm{T}$ be less than 0.1 if $E_\mathrm{T}>$20 GeV.
Electrons arising from photons that convert into $e^+e^-$ pairs are removed by cuts
applied to pairs of opposite-charge tracks with small opening angles.
The electron identification efficiency was measured with Drell-Yan (DY) electrons to range
from 75 to $83\%$ \cite{id_effic} increasing with the electron $E_\mathrm{T}$.

Muons are identified by matching tracks with reconstructed track stubs in the
CMUP or CMX. The energy deposited in the calorimeters must be
consistent with a minimum-ionizing particle. The isolation requirement is similar
to that for electrons:  $E_\mathrm{T}^\mathrm{iso}<2$ GeV if $p_\mathrm{T}\le 20$ GeV/$c$ or
$E_\mathrm{T}^\mathrm{iso}/E_\mathrm{T}<0.1$ if $p_\mathrm{T}>20$ GeV/$c$.
The muon identification efficiency was measured with $J/\psi$ and $Z$ data to
be in the range $90-96\%$ \cite{id_effic}
rising with the muon  $p_\mathrm{T}$.

For both electrons and muons the impact parameter ($d_0$) of the track with respect to
the beam line position must be less than 0.2 cm if the track is based on COT information
only or less than 0.02 cm if silicon-based tracking is also available.
The longitudinal position of the
event vertex is required to be within 60 cm of the center of the detector.

Jets are identified in $\eta-\phi$ space as a group of electromagnetic and hadronic
towers using a clustering algorithm with cone size
$\Delta R=\sqrt{(\Delta\eta)^2+(\Delta\phi)^2} = 0.7$.
The jet energy resolution
is $\sigma \simeq  0.1 \cdot  E_{\mathrm{T}}~{\rm (GeV)} \oplus 1$ GeV for jets with
$35 < E_\mathrm{T} < 450$ GeV.
In order to find genuine hadronic jets and to
avoid counting electrons and photons as jets, the
fraction of the total energy deposited in the electromagnetic calorimeters is
required to be between 0.1 and 0.9. Corrections to the jet energy are applied to take
into account $\eta$-dependent losses, luminosity-dependent multiple interactions,
and non-linearities in calorimeter response \cite{jes}.
In this analysis we require at least one jet with $|\eta|<2$
and $E_\mathrm{T}>15$ GeV after corrections.

$\not$$E_\mathrm{T}$  is calculated as the magnitude of the
negative vector sum of the transverse energy
deposited in the calorimeter towers with $|\eta|<3.6$ and energy larger
than 0.1 GeV. Since
muons are minimum-ionizing particles and deposit only $\simeq2$ GeV
energy in the calorimetry, the muon momentum from
tracking (minus the expected energy deposition) is used to
correct $\not$$E_\mathrm{T}$. The energy corrections applied to jets with
$|\eta|<2.4$ and $E_\mathrm{T}>10$ GeV  are also propagated to the calculation of $\not$$E_\mathrm{T}$.

\section{BACKGROUND ESTIMATION}

Several background sources result in events with dileptons,
jets, and $\not$$E_\mathrm{T}$.
These backgrounds are $t \bar t$  production, other heavy-flavor
quark ($b \bar b$ and $c \bar c$) production,
DY production of lepton pairs
where mismeasurement of a jet or lepton results in substantial $\not$$E_\mathrm{T}$,
diboson production ($WW$, $WZ$, $ZZ$, and $W\gamma$), and events with a
lepton and a misidentified or secondary lepton ($l$+fake).
The jets in most backgrounds result from QCD initial-state or final-state
radiation (ISR or FSR).

The $t \bar t$, DY, and diboson backgrounds are estimated by generating
Monte Carlo~(MC) events using  the {\sc pythia} \cite{pythia} event generator
followed by a  {\sc geant}-based \cite{geant}
detector simulation. Backgrounds arising from fake leptons and heavy
flavor ($b \bar b$, $c \bar c$) are estimated using data-driven methods.

The MC $t \bar t$  samples are normalized to the next-to-leading-order (NLO) cross section
$\sigma_{t \bar t}=6.71$ pb \cite{ttbar} with a top mass of 175 GeV/$c^2$. The $Z\rightarrow ee$,
$Z\rightarrow \mu\mu$, and
$Z\rightarrow \tau\tau$ samples are normalized to a leading-order cross section times
branching ratio of 1272 pb for $m_{ll} > 10$ GeV/$c^2$ times a $k$-factor of 1.4 \cite{k-factor}
to correct for NLO contributions.
The NLO cross sections times branching ratios for $WW \rightarrow ll\nu\nu$,
$WZ \rightarrow llX$, and $ZZ \rightarrow llX$ are  1.27, 0.365,
and 1.513 pb  respectively \cite{WW}.
For $W\gamma\rightarrow e\nu\gamma$ and  $W\gamma\rightarrow \mu\nu\gamma$,
we use the leading order cross section of 21.5 pb times a $k$-factor of 1.34 to
account for NLO contributions \cite{baur}.

The lepton plus fake background consists of events with a genuine lepton
plus a ``fake'' lepton which is either a light hadron misidentified as a lepton
or an uninteresting lepton from pion or kaon decay-in-flight.
For muons the fakes can be particles that penetrate the calorimeters and absorbing
material and reach the muon detectors or decay in flight to muons. In the case of
electrons, fakes are  usually jets that are misidentified as electrons, mainly due to
neutral pions that decay to photons which shower in the electromagnetic calorimeters.
This background is
estimated by examining samples of single-lepton events taken with the
$E_\mathrm{T}>8$~GeV electron calibration trigger and the $p_\mathrm{T}>8$~GeV/$c$
muon calibration trigger. Events with at least one $E_\mathrm{T}>4$ GeV central jet ($|\eta|<1.1$)
or one isolated track with  $p_\mathrm{T}>4$ GeV/$c$
passing the muon track quality cuts, excluding the trigger lepton,
form the $l$+fake-electron and $l$+fake-muon
candidate samples, respectively. To determine the background, it is necessary to estimate
a fake rate from
samples of jet events that contain a  negligible number of directly produced  leptons.
Four samples containing jets with $E_\mathrm{T} > 20$, 50, 70, and 100 GeV, respectively, are used.
The electron fake rate is defined as the probability of a jet being misidentified as an
electron and the muon fake rate as the probability of an isolated track being misidentified
as a muon.
The fake rates are determined as a function of jet  $E_\mathrm{T}$ for electrons
and of track $ p_\mathrm{T}$ for muons. The fake rate for 20 GeV electrons is about
0.0002 fakes per jet and the fake rate for 20 GeV/$c$ muons is about 0.004 fakes per track.
These fake rates are then applied to each fake-electron and fake-muon candidate, one
at a time, in the single lepton sample.
Events that pass the analysis cuts are assigned a weight,
which is the appropriate fake rate, scaled to the integrated
luminosity of the dilepton data sample relative to the integrated luminosity
of the single lepton sample, taking into account the single lepton trigger prescale.
The sum of the weights then forms the $l$+fake background.
A 50\% systematic uncertainty on the misidentifed lepton background is
assigned based upon the differences between the four jet samples used to determine
the fake rates.

The background arising from heavy-flavor ($b \bar b$, $c \bar c$) is estimated
utilizing dilepton events enriched in heavy-flavor events by inverting the
normal impact parameter requirements, i.e.
$|d_0| > 0.2$~cm for COT-only tracks and
$|d_0| > 0.02$~cm if the track included information from the SVX, and
requiring that at least one lepton pass the inverse $d_0$ cuts.
A ``scaling region''  is defined as  the dilepton invariant mass
range 15 $<$ $M_{ll}$ $<$ 35~GeV/$c^2$.
The scaling region has no requirements imposed on $\not$$E_\mathrm{T}$, the number of jets,
or other kinematic variables.
In this  region the only significant contributions
to the data sample with normal $d_0$ cuts
are due to DY, heavy-flavor, and
light hadrons misidentified as leptons. The DY contribution is derived from
MC samples, and the misidentified lepton component is
estimated using the technique described above.
The remaining events
are attributed to heavy-flavor (HF) and are used to calculate scaling factors,
defined as the ratio of these inferred HF events passing the normal $d_0$ cuts
to the total number of events  passing the above ``inverse $d_0$'' cuts.
It is assumed that this scaling ratio is independent
of dilepton mass in regions not dominated by $Z$ boson decays.
To estimate the heavy-flavor background,
the resulting scaling factors, typically of order 2,
are then applied to ``inverse $d_0$'' events passing cuts appropriate to the
various control regions as well as the signal region.
No heavy-flavor contribution to the signal region survives our final cuts.

\begin{table}[htpb]
\begin{center}
\begin{ruledtabular}
\begin{tabular}{cccc}
 & $ee$ &  $e\mu$ &  $\mu\mu$  \\
\hline
DY & 72.8$\pm$4.8$\pm$26.3 & 26.6$\pm$2.7$\pm$5.4 & 62.4$\pm$4.1$\pm$28.4  \\
\ttbar & 6.1$\pm$0.1$\pm$0.7 & 13.1$\pm$0.1$\pm$1.4 & 4.2$\pm$0.1$\pm$0.5  \\
diboson & 3.5$\pm$0.1$\pm$0.6 & 6.2$\pm$0.1$\pm$1.1 & 2.1$\pm$0.0$\pm$0.4  \\
$l$+fake & 21.6$\pm$0.2$\pm$10.8 & 24.9$\pm$0.4$\pm$12.4 & 5.4$\pm$0.2$\pm$2.7  \\
HF & 9.1$\pm$4.1$\pm$7.4 & 30.6$\pm$7.9$\pm$10.5 & 8.5$\pm$4.3$\pm$6.7  \\
\hline
Total & 113$\pm$6$\pm$30 & 101$\pm$8$\pm$18 & 83$\pm$6$\pm$30  \\
\hline
Data & 110 & 76 & 89  \\
\end{tabular}
\end{ruledtabular}
\end{center}
\hfill{}
\caption{ Expected backgrounds (DY, $t \bar t$, diboson, misidentified hadrons
and decays-in-flight ($l$+fake), and $b \bar b$ and $c \bar c$ (HF))
and observed events in the pre-signal region. The first uncertainty listed is statistical and the
second systematic. The systematic uncertainty includes a $6\%$ uncertainty on the luminosity
common to all entries.}
\label{crSt}
\end{table}

\begin{table}[htpb]
\begin{center}
\begin{ruledtabular}
\begin{tabular}{ccc}
 & $ee$ &  $\mu\mu$ \\
\hline
DY & 12314$\pm$63$\pm$956 & 6904$\pm$43$\pm$568 \\
\hline
Data & 12461 & 7111 \\
\end{tabular}
\end{ruledtabular}
\end{center}
\caption{ Expected DY background
and observed events in the $Z$ control region. The first uncertainty listed is statistical and the
second systematic. The systematic uncertainty includes a $6\%$ uncertainty on the luminosity
common to both channels.}
\label{crAt}
\end{table}

\begin{table}[htpb]
\begin{center}
\begin{ruledtabular}
\begin{tabular}{cccc}
 & $ee$ &  $e\mu$ &  $\mu\mu$ \\
\hline
DY & 167$\pm$7$\pm$26 & 94$\pm$5$\pm$7 & 108$\pm$5$\pm$19 \\
\ttbar & 0.1$\pm$0.0$\pm$0.0 & 0.2$\pm$0.0$\pm$0.0 & - \\
diboson & 9.3$\pm$0.0$\pm$0.9 & 17.9$\pm$0.1$\pm$1.7 & 5.9$\pm$0.0$\pm$0.6 \\
$e$+fake & 23.1$\pm$0.1$\pm$11.5 & 11.8$\pm$0.3$\pm$5.9 & - \\
\mufk & - & 14.8$\pm$0.1$\pm$7.4 & 4.7$\pm$0.2$\pm$2.4 \\
HF & 15.6$\pm$5.3$\pm$9.6 & 62.6$\pm$11.9$\pm$3.8 & 26.2$\pm$8.1$\pm$6.7 \\
\hline
Total & 215$\pm$9$\pm$30 & 202$\pm$13$\pm$13 & 145$\pm$10$\pm$20 \\
\hline
& & & \\
Data & 186 & 167 & 114 \\
\end{tabular}
\end{ruledtabular}
\end{center}
\caption{ Expected backgrounds (DY, $t \bar t$, diboson, misidentified hadrons
and decays-in-flight ($e$+fake, $\mu$+fake), and $b \bar b$ and $c \bar c$ (HF))
and observed events for the high-$\not$$E_\mathrm{T}$ / no jet control region.
The first uncertainty listed is statistical and the second systematic.
The systematic uncertainty includes a $6\%$ uncertainty on the luminosity common to
all entries.}
\label{crDt}
\end{table}

\begin{table}[htpb]
\begin{center}
\begin{ruledtabular}
\begin{tabular}{cccc}
 & $ee$ &  $e\mu$ &  $\mu\mu$ \\
\hline
DY & 6254$\pm$43$\pm$644 & 201$\pm$8$\pm$21 & 4681$\pm$35$\pm$526 \\
\ttbar & - & - & - \\
diboson & 1.4$\pm$0.0$\pm$0.1 & 2.5$\pm$0.0$\pm$0.2 & 0.8$\pm$0.0$\pm$0.1 \\
$e$+fake & 263$\pm$0$\pm$132 & 95$\pm$1$\pm$48 & - \\
\mufk & - & 105$\pm$0$\pm$52 & 40$\pm$0$\pm$20 \\
HF & 826$\pm$86$\pm$138 & 1102$\pm$72$\pm$110 & 554$\pm$71$\pm$66 \\
\hline
Total & 7345$\pm$96$\pm$671 & 1505$\pm$72$\pm$132 & 5276$\pm$80$\pm$530 \\
\hline
& & & \\
Data & 7448 & 1687 & 5344 \\
\end{tabular}
\end{ruledtabular}
\end{center}
\caption{ Expected backgrounds (DY, $t \bar t$, diboson, misidentified hadrons
and decays-in-flight ($e$+fake, $\mu$+fake), and $b \bar b$ and $c \bar c$ (HF))
and observed events for the low-$\not$$E_\mathrm{T}$ / no jet control region.
The first uncertainty listed is statistical and the second systematic.
The systematic uncertainty includes a $6\%$ uncertainty on the luminosity common to all
entries.}
\label{crB0C0}
\end{table}

\begin{table}[htpb]
\begin{center}
\begin{ruledtabular}
\begin{tabular}{cccc}
 & $ee$ &  $e\mu$ &  $\mu\mu$ \\
\hline
DY & 1161$\pm$19$\pm$223 & 42$\pm$3$\pm$6 & 1004$\pm$16$\pm$204 \\
\ttbar & 0.3$\pm$0.0$\pm$0.0 & 0.6$\pm$0.0$\pm$0.1 & 0.2$\pm$0.0$\pm$0.0 \\
diboson & 1.4$\pm$0.0$\pm$0.1 & 0.6$\pm$0.0$\pm$0.1 & 1.0$\pm$0.0$\pm$0.1 \\
$e$+fake & 215$\pm$0$\pm$107 & 61$\pm$1$\pm$31 & - \\
\mufk & - & 94$\pm$0$\pm$47 & 28$\pm$0$\pm$14 \\
HF & 144$\pm$21$\pm$32 & 247$\pm$26$\pm$70 & 170$\pm$27$\pm$46 \\
\hline
Total & 1521$\pm$28$\pm$250 & 445$\pm$26$\pm$90 & 1204$\pm$32$\pm$209 \\
\hline
& & & \\
Data & 1443 & 351 & 1246 \\
\end{tabular}
\end{ruledtabular}
\end{center}
\caption{ Expected backgrounds (DY, $t \bar t$, diboson, misidentified hadrons
and decays-in-flight ($e$+fake, $\mu$+fake), and $b \bar b$ and $c \bar c$ (HF))
and observed events for the low-$\not$$E_\mathrm{T}$ / one or more jet control region.
The first uncertainty listed is statistical and the second systematic.
The systematic uncertainty includes a $6\%$ uncertainty on the luminosity common to
all entries.}
\label{crB1C1}
\end{table}

\begin{table}[htpb]
\begin{center}
\begin{ruledtabular}
\begin{tabular}{lcccc}
 \multicolumn{1}{c} {Variable} & \multicolumn{4}{c} { Cut group } \\
& $a$ & $b$ & $c$ & $d$   \\
\hline
$\Delta m $(GeV/$c^2$) & $5-47.5$ & $47.5-72.5$ &  $72.5-87.5$ &  $>87.5$  \\
\hline
$\not$$E_\mathrm{T}$  (GeV) & $>$25 & $>$32 & $>$32 & $>$32 \\
$\Delta \phi(p_\mathrm{T}^{ll}$,$\not$$E_\mathrm{T}$) (deg) & $>60$ & $>60$ & $>60$ & $>60$ \\
$p_\mathrm{T2}$ (GeV/$c$)  & $>$7 & $>$7 & $>$7 & $>$7 \\
$H_\mathrm{T}$ (GeV) & $<170$ &  $120-225$ &  $130-290$ &  $>165$ \\
$m_\mathrm{T}$ (GeV/$c^2$) & $>$15 & $>$11 & --- & --- \\
$p_\mathrm{T}^{ll}$ (GeV/$c$) & $< c\Delta m - 1$ & --- & --- & --- \\
\end{tabular}
\end{ruledtabular}
\caption{ \label{CutGrt} Table of cuts for the cut groups defining the four signal regions.}
\end{center}
\end{table}

\section{SIGNAL SAMPLES AND SYSTEMATIC UNCERTAINTIES}

A total of  74 MC signal samples corresponding to values
of $m_{\tilde{t}}$ = [55,190] GeV/$c^2$
and $m_{\tilde{\nu}}$ = [45,110] GeV/$c^2$ were generated
using {\sc pythia} and the {\sc geant}-based
detector simulation. The events are scaled using
the NLO {\sc prospino} cross section \cite{Prospino1,prospino}
calculated with the CTEQ6M \cite{cteq} parton distribution functions (PDFs).
The cross sections depend on the stop quark mass and only weakly,
through higher-order corrections, on other SUSY parameters. Limits on the production
cross section can therefore be translated into lower limits on the lightest stop quark
mass without reference to other SUSY parameters.
The branching ratio for the
decay  $\Stop \rightarrow b l^+ \tilde{\nu_l}$, where $l =  e$, $\mu$ or $\tau$ with
equal probablity, is assumed to be $100\%$.

There are several sources of systematic uncertainty on the background estimation.
The systematic uncertainty due to the jet energy scale
is determined by varying the  jet energy
corrections by $\pm 1 \sigma$ \cite{jes}.
The resulting uncertainty varies from less than $1\%$ to $35\%$ and is largest
for DY events, which typically contain low-$E_\mathrm{T}$ jets.
The uncertainty arising from ISR and FSR
is determined by varying the parameters in {\sc pythia} that control the generation of ISR/FSR.
The resulting uncertainty is  $\simeq3\%$.
The uncertainty on the acceptance arising
from the PDFs used in the MC is estimated using the uncertainties on the CTEQ eigenvectors
and determined to be $2\%$.
Other systematic uncertainties are: $6\%$ on the measurement of the integrated luminosity,
$2\%$ on the dilepton trigger efficiency, $2\%$ on lepton identification efficiency,
and $50\%$ on the number of misidentified electrons and muons.
The uncertainties on the cross sections used in the MC generation of the background
are  $8\%$ for $t\bar{t}$, $2\%$ for
DY, $6\%$ for $WW$, $8\%$ for $WZ$, $10\%$ for $ZZ$, and $7\%$ for $W\gamma$.

The systematic uncertainties on the MC stop signal estimation are nearly identical to those on the
background estimation. The uncertainty arising from the jet energy scale varies from $1\%$ to
$11\%$ depending on the stop-sneutrino mass difference. The uncertainties due to ISR/FSR,
PDFs, luminosity, trigger efficiency, and lepton identification are the same as for the
background estimation.

\section{INITIAL EVENT SELECTION AND CONTROL SAMPLES}

We first define a pre-signal region by applying several
event selection cuts to
significantly reduce background and provide
a data sample loosely consistent
with the stop quark signature.
We subsequently optimize additional cuts to improve
the sensitivity of the search. This is done prior to revealing the contents
of the data in the pre-signal region.
At the pre-signal stage  the following cut requirements are applied:
two opposite-charge leptons, one with
\pT $> 10$~GeV/$c$ and the other with  \pT $> 5$~GeV/$c$;
$m_{ll} > 15$~GeV/$c^2$, in order
to remove sequential $B$-hadron decays and low mass resonances;
$m_{ll} < 76$~GeV/$c^2$ or $m_{ll} > 106$~GeV/$c^2$ for
same-flavor dilepton events,
in order to eliminate $Z$ boson events;
at least one jet with corrected $E_\mathrm{T} > 15$~GeV and
$|\eta| < 2$;
$\not$$E_\mathrm{T}$ $>$ 15~GeV; $\Delta R$($e$,highest-$E_\mathrm{T}$ jet) $>$ 0.4;
$\Delta R$($l,l$) $>$ 0.4; and $\Delta\phi > 20^{\circ}$
between $\not$$E_\mathrm{T}$ and
each of the leptons and the hightest $E_\mathrm{T}$ jet.

Figure~\ref{fig:MET_presig} shows the $\not$$E_\mathrm{T}$ distributions separately
for $ee$,
$e\mu$ and $\mu\mu$ events, as well as the summed distribution, in the
pre-signal region. The expected $\not$$E_\mathrm{T}$ distribution for stop quark events
with stop mass 150 GeV/$c^2$ and sneutrino mass 75 GeV/$c^2$ is also
shown scaled up by a factor of five.
Figures~\ref{fig:HT_presig}-\ref{fig:PT2_presig} show the corresponding
plots for $H_\mathrm{T}$,
the $p_\mathrm{T}$ of the highest $p_\mathrm{T}$ lepton ($p_\mathrm{T1}$), and the $p_\mathrm{T}$
of the next-to-highest $p_\mathrm{T}$ lepton ($p_\mathrm{T2}$).
$H_\mathrm{T}$ is defined
as $H_\mathrm{T}$ = $\not$$E_\mathrm{T}$+$p_\mathrm{T1}$+$p_\mathrm{T2}$+$E_\mathrm{Tj}$,
where $E_\mathrm{Tj}$ is the transverse energy of the
highest $p_\mathrm{T}$ jet.
Table~\ref{crSt} lists the sources of expected background for the pre-signal
region and the number of observed events.
Good agreement of data
with the background estimations is observed.

To check the accuracy of our estimation of  SM backgrounds,
a number of ``control regions'' are defined.
One control region consists of same-flavor, opposite-charge lepton events with
invariant mass
76 $<$ $m_{ll}$ $<$ 106~GeV/$c^2$ (the $Z$ region). No jet or $\not$$E_\mathrm{T}$ requirements
are imposed.
The background is expected to arise almost entirely from DY processes.
Table~\ref{crAt} gives the number of observed events in the $Z$ region and the expected
background. Good agreement is seen, demonstrating accurate modeling of the predominantly
DY background.

Another control region consists of events where we require  $\not$$E_\mathrm{T}$ $>$ 15~GeV and no jets
with \ET $> 15$ GeV. All the other pre-signal cuts are applied
except for the  $\Delta\phi > 20^{\circ}$ cut between $\not$$E_\mathrm{T}$ and
each of the two highest $p_\mathrm{T}$ leptons and the highest $E_\mathrm{T}$ jet. This control region
predominantly tests the modeling of the high $\not$$E_\mathrm{T}$ tail of DY and HF events.
Table~\ref{crDt} gives a breakdown of the expected backgrounds and the number of observed events
for this control region.
Reasonable agreement is observed between the expected background and data. The significant
number of $e\mu$ DY events (comparable to the number of $ee$ and $\mu\mu$ events)
is due to $Z\rightarrow \tau\tau$ decays where one $\tau$ subsequently decays to an
electron and the other to a muon.

Two additional control regions are defined as having $\not$$E_\mathrm{T} < 15$ GeV
and either the presence or absence of jets with $E_\mathrm{T} > 15$~GeV.
These control regions are sensitive to the modeling of heavy flavor and the photon
component of DY.
All other cuts are the same as for the previous control region.
Tables~\ref{crB0C0} and \ref{crB1C1}
list the expected backgrounds and the number of observed events
for these control regions. In both cases, the data agree well with the expected backgrounds.

\begin{table*}[htpb]
\begin{center}
\begin{ruledtabular}
\begin{tabular}{ccccccccc}
cut group & flavor  & DY &  \ttbar &  diboson  & $l$+fake & Total Background  & $\Stop\antiStop$ & Data \\
\hline
$a$ & $ee$ & 0.9$\pm$0.4 & 0.1$\pm$0.0 & 0.5$\pm$0.1 & 1.7$\pm$0.9 & 3.3$\pm$0.9 & 3.2$\pm$0.5 & 1 \\
& $e\mu$ & 0.3$\pm$0.1 & 0.3$\pm$0.0 & 1.0$\pm$0.2 & 2.0$\pm$1.0 & 3.6$\pm$1.0 & 6.6$\pm$0.8 & 2 \\
& $\mu\mu$ & 0.6$\pm$0.4  & 0.1$\pm$0.0 & 0.3$\pm$0.1 & 0.2$\pm$0.1 &  1.1$\pm$0.4 & 2.9$\pm$0.4 & 1 \\
& $ee+e\mu+\mu\mu$ & 1.8$\pm$0.7 & 0.5$\pm$0.1 & 1.8$\pm$0.3 & 3.9$\pm$1.7 & 7.9$\pm$1.9 & 12.8$\pm$1.4 & 4 \\

\hline
$b$ & $ee$ & 0.5$\pm$0.3 & 1.6$\pm$0.2 & 1.3$\pm$0.2 & 1.6$\pm$0.8 & 4.9$\pm$0.9 & 3.7$\pm$0.4 & 3 \\
& $e\mu$ & 0.3$\pm$0.1 & 3.2$\pm$0.3 & 2.8$\pm$0.5 & 2.5$\pm$1.3 & 8.7$\pm$1.4 & 8.8$\pm$0.8 & 11 \\
& $\mu\mu$ & 0.3$\pm$0.1  & 1.1$\pm$0.1 & 1.0$\pm$0.2 & 0.2$\pm$0.1 & 2.4$\pm$0.3 & 3.0$\pm$0.4 & 8 \\
& $ee+e\mu+\mu\mu$ & 1.0$\pm$0.4 & 5.8$\pm$0.6 & 5.0$\pm$0.9 & 4.3$\pm$1.8 & 16.1$\pm$2.3 & 15.6$\pm$1.4 & 22 \\
\hline
$c$ & $ee$ & 0.8$\pm$0.4 & 3.3$\pm$0.4 & 1.4$\pm$0.2 & 1.3$\pm$0.6 & 6.7$\pm$0.8 & 6.1$\pm$0.5 & 7 \\
& $e\mu$ & 0.3$\pm$0.1 & 7.1$\pm$0.8 & 2.9$\pm$0.5 & 2.1$\pm$1.1 & 12.5$\pm$1.4 & 12.6$\pm$0.9 & 13 \\
& $\mu\mu$ & 0.4$\pm$0.2  & 2.3$\pm$0.3 & 1.0$\pm$0.2 & 0.2$\pm$0.1 & 3.9$\pm$0.4 & 4.2$\pm$0.4 & 9 \\
& $ee+e\mu+\mu\mu$ & 1.5$\pm$0.5 & 12.7$\pm$1.4 & 5.3$\pm$0.9 & 3.6$\pm$1.5 & 23.1$\pm$2.6 & 22.9$\pm$1.6 & 29 \\
\hline
$d$ & $ee$ & 0.4$\pm$0.1 & 4.4$\pm$0.5 & 0.9$\pm$0.2 & 0.6$\pm$0.3 & 6.2$\pm$0.6 & 3.4$\pm$0.3 & 5 \\
& $e\mu$ & 0.1$\pm$0.0 & 9.4$\pm$1.0 & 1.9$\pm$0.3 & 1.4$\pm$0.7 & 12.7$\pm$1.3 & 7.3$\pm$0.5 &  11 \\
& $\mu\mu$ & 0.5$\pm$0.2  & 3.0$\pm$0.4 & 0.7$\pm$0.1 & 0.2$\pm$0.1 &4.4$\pm$0.5  & 2.3$\pm$0.2 & 8 \\
& $ee+e\mu+\mu\mu$ & 1.0$\pm$0.4 & 16.8$\pm$1.8 & 3.5$\pm$0.6 & 2.1$\pm$0.9 & 23.3$\pm$2.4 & 13.0$\pm$0.9 & 24 \\
\end{tabular}
\end{ruledtabular}
\caption{ \label{cgdt} Expected backgrounds and observed events for cut groups $a$ - $d$.
The uncertainties represent the statistical and systematic uncertainties added in quadrature.
The systematic uncertainty includes a $6\%$ uncertainty on the luminosity common to
all entries.
Also shown are the number of expected $\Stop\antiStop$ events for (stop, sneutrino) masses
(130,95), (140,90), (150,75), and (180,50) GeV/$c^2$ in cut groups $a,b,c$ and $d$ respectively.}
\end{center}
\end{table*}

\section{OPTIMIZATION OF CUTS}

The signal regions are determined by optimizing cuts on variables that
minimize the limit on the $\Stop\antiStop$ production cross section determined with
Bayesian methods \cite{Bayes}.
The $ee, e\mu$, and $\mu\mu$ final states
are tuned simultaneously and flavor-independent cut values are obtained.
The most useful variables for discriminating  stop quark signal
from background are: $\not$$E_\mathrm{T}$; $H_\mathrm{T}$;
the absolute value of the
difference in azimuthal angle between the dilepton system and  $\not$$E_\mathrm{T}$; $p_\mathrm{T2}$;
the transverse momentum of the dilepton system ($p_\mathrm{T}^{ll}$); and
the transverse mass between each lepton (l) and $\not$$E_\mathrm{T}$, defined as
$m_\mathrm{Ti} = \sqrt{2p_\mathrm{Ti}\MET[1 - \cos(\Delta\phi(l_i,\MET ))]}$
where the index i labels the lepton.

The optimum values for the cut variables depend on the
mass difference ($\Delta m = m_{\tilde{t}} - m_{\tilde{\nu}}$)
which we group into
four sets labeled {$a$} through {$d$} in $\Delta m$  bands parallel to and below the
{$m_{\tilde{t}} = m_{\tilde{\nu}} + m_{b}$} kinematic limit.
Four cut groups were chosen as a compromise between having the cuts near their optimal
values for each stop-sneutrino point while keeping the number of cut groups to a minimum.
The definitions of the four cut groups and the values of the cuts used
are given in Table~\ref{CutGrt}.

The $\not$$E_\mathrm{T}$ distributions for the signal region are shown in
Fig.~\ref{fig:MET_sigb} for cut group $b$, where the other cuts listed in Table~\ref{CutGrt}
are all applied. The individual backgrounds are shown as well as the data.
For reference the expected signal from the stop-sneutrino mass point (140,90) GeV/$c^2$
is also shown added to the stacked backgrounds.
The vertical line represents the lower bound placed on $\not$$E_\mathrm{T}$ for this cut group.
The $\not$$E_\mathrm{T}$ cut is the most effective at reducing the DY, $b \bar b$ and  $c \bar c$, and
$l$+fake backgrounds.
In particular, the DY background is expected to show a small energy imbalance
in the detector,
while the reference-point signal events are characterized by large energy imbalance.
The cut on $\not$$E_\mathrm{T}$ reduces the
DY background by more than a factor of 10
even when all other final cuts have been
applied, while reducing the expected  signal by only $20\%$.

As shown in Fig.~\ref{fig:phi12M_sigb},
the cut on the absolute value of the
azimuthal angle between the dilepton system and $\not$$E_\mathrm{T}$ is useful
in further supressing the dominant DY background. This cut discriminates
against DY events where the $\not$$E_\mathrm{T}$ arises from mismeasurement of
the leptons or from $\tau\tau$ events where neutrinos from the $\tau$ decay result in
$\not$$E_\mathrm{T}$ aligned with the leptons coming from the decay.

Leptons arising from stop quark decay are typically more energetic than those coming
from $b$ or $c$ quark decay because of the higher stop quark mass.
A cut on $p_\mathrm{T2}$, whose distribution is shown in Fig.~\ref{fig:pt2_sigb},
removes all the $b \bar b$ and $c \bar c$
background remaining after the other cuts have been applied. Only $5\%$ of the
reference-point signal events are removed by this cut.

The $t \bar t$ background is especially difficult to reduce without severely impacting the
efficiency for stop quark detection. However, because the
sneutrino carries a significant fraction of the available energy, the remaining
stop quark decay products typically have less energy than their counterparts from
top quark decay. As a result the $H_\mathrm{T}$ distribution, as shown in Fig.~\ref{fig:Ht_sigb},
peaks at lower values for stop quarks than for top quarks.
An upper limit on $H_\mathrm{T}$ is effective at reducing the amount of $t \bar t$ background while
a lower limit helps reduce the other backgrounds.

A transverse mass cut is
used in the  low $\Delta m$ cut groups ($a$ and $b$)
to remove most of the DY background remaining after the other cuts have
been applied. The transverse mass distribution for cut group $b$ is shown in
Fig.~\ref{fig:Mt_sigb}.

The upper bound on the transverse momentum of the dilepton system is applied to the
lowest   $\Delta m$ cut group ($a$) to reduce the $t \bar t$ and diboson backgrounds. Its
effectiveness  decreases as  $\Delta m$ increases.
The sliding cut is determined by a linear fit to the optimal dilepton $p_\mathrm{T}$
values as a function of $\Delta m$ in cut group $a$: $p_\mathrm{T}^{ll} < (c\Delta m - 1)$ GeV/$c$.

\section{RESULTS AND CONCLUSIONS}

The numbers of expected background and observed data events for cut groups $a-d$
are given in Table~\ref{cgdt}.
In general the agreement between data and SM background estimations is good,
although the statistical uncertainties are large. The largest deviation is an excess of
data over background expectations for $\mu\mu$ events with large $\not$$E_\mathrm{T}$.
Examination of individual event properties found no evidence
for cosmic rays or
pion or kaon decays-in-flight.
These results are not independent observations, since there
is a large overlap in events between cut groups $b$, $c$, and $d$.
The largest muon excess, found in cut group $b$, has a 0.4\% probability for
the modeled background to fluctuate up to the number of observed events or more (p-value),
when considering this cut group alone.
The p-value includes the effects of the estimated background uncertainty.
Combining all channels in cut group $b$ raises the probability to 12.4\%.
The corresponding one-sided Gaussian significances are 2.6~$\sigma$ and
1.2~$\sigma$ respectively.

A joint likelihood is formed from the product of the individual channel likelihoods.
Using this likelihood, we apply a Bayesian method \cite{Bayes} with a flat prior
for the signal to set 95\% confidence level upper limits on the production
cross section at each considered point in the stop-sneutrino mass plane.
Systematic uncertainties are incorporated by convolving the Poisson probabilities
for the signal with Gaussian distributions representing each background uncertainty.
The three dilepton flavor channels are incorporated into the statistical
analysis simultaneously (but individually), with full treatment of correlated
and uncorrelated uncertainties. The correlations between the various backgrounds
and the MC-generated signal are also accounted for.

One-dimensional curves of the upper cross section limits and the theoretical
cross section are shown in Fig.~\ref{fig:xs2} for groups of points with fixed $\Delta m$.
The cross section upper limits for a given $\Delta m$ tend to
be almost independent of the stop mass.

To determine the observed (expected) exclusion contour in the stop-sneutrino mass plane,
we set the number of events equal to the number of events in the data (total expected
background).
We then calculate the 95\% confidence level upper limits
on the stop pair cross section for 74 points in the $(m_{\tilde{t}},m_{\tilde{\nu}})$
plane to determine both the points we exclude and those that we
expect to exclude in the absence of signal.
We interpolate linearly between nearby excluded and not-excluded points.
The expected and observed limits are shown in Fig.~\ref{fig:obsL} along with
previous limits \cite{LEP, CDFD0}.

In conclusion,
this analysis extends the previously existing exclusion limits to higher sneutrino masses for stop masses in
the range 135-155~GeV/$c^2$, and to stop masses up to 180~GeV/$c^2$, for low
sneutrino masses.
For a particular set of optimization cuts (cut group $b$),
a 2.6~$\sigma$ excess
of $\mu\mu$ events is observed, but is reduced to 1.2~$\sigma$ when all channels are combined.

\section{ACKNOWLEDGMENTS}

We thank the Fermilab staff and the technical staffs of the participating institutions for their vital contributions.
This work was supported by the U.S. Department of Energy and National Science Foundation; the Italian Istituto
Nazionale di Fisica Nucleare; the Ministry of Education, Culture, Sports, Science and Technology of Japan; the
Natural Sciences and Engineering Research Council of Canada; the National Science Council of the Republic of China;
the Swiss National Science Foundation; the A.P. Sloan Foundation; the Bundesministerium f\"ur Bildung und Forschung,
Germany; the World Class University Program, the National Research Foundation of Korea; the Science and Technology
Facilities Council and the Royal Society, UK; the Institut National de Physique Nucleaire et Physique des
Particules/CNRS; the Russian Foundation for Basic Research; the Ministerio de Ciencia e Innovaci\'{o}n, and Programa
Consolider-Ingenio 2010, Spain; the Slovak R\&D Agency; and the Academy of Finland.

\nopagebreak


\begin{thebibliography}{99}
\bibitem{met}
CDF uses a cylindrical coordinate system in which $\phi$
is the azimuthal angle, $r$ is the radius from the nominal beam line, and
$z$ points in the proton beam direction. The transverse ($r-\phi$) plane is
perpendicular to the $z$ axis. Transverse momentum and energy are the respective
projections of momentum measured in the tracking system and energy measured in the
calorimeter system onto the $r-\phi$ plane, and are defined as
$p_\mathrm{T}$ = $p\sin\theta$ and $E_\mathrm{T}$ = $E\sin\theta$. Here, $\theta$ is the polar angle
measured with respect to the interaction vertex. Missing transverse
energy ($\overrightarrow{\not$$E_\mathrm{T}}$)
is defined
by $\overrightarrow{\not$$E_\mathrm{T}}$ = $-\Sigma_{i}E_\mathrm{T}^{i}\overrightarrow{n_{i}}$,
where $i$ is the calorimeter tower
and $\overrightarrow{n}_{i}$ is a unit vector perpendicular to the
beam axis and pointing at the $i$th tower. The pseudorapidity $\eta$ is defined as
-ln(tan($\theta/2$)), where $\theta$ is measured with respect to the origin of the
detector. We define the magnitude $\not$$E_\mathrm{T}$~=~$|\overrightarrow{\not$$E_\mathrm{T}}|$.
\end{thebibliography}
\end{document}